\documentclass[11pt, a4paper]{article}

\usepackage{booktabs}
\usepackage[toc,page]{appendix}
\usepackage{amsmath}
\usepackage{amsfonts}
\usepackage{amssymb}
\usepackage{graphicx, rotating}
\usepackage{epstopdf}
\usepackage{epsfig}
\usepackage{braket}
\usepackage{simplewick}
\usepackage{bbm}
\usepackage{empheq}
\usepackage{latexsym}
\usepackage{graphicx}
\usepackage{leftidx}
\usepackage{subcaption}
\usepackage{color}
\usepackage{amsmath,amssymb}
\usepackage{cite}
\usepackage{slashed}
\usepackage{hyperref}
\usepackage{xcolor}
\usepackage{bm}

\usepackage{tikz}
\usepackage[compat=1.1.0]{tikz-feynman}

\newcommand{\deltazero}{
\begin{tikzpicture}[baseline]
\begin{feynman}
\vertex (c);
\vertex [above=0.6cm of c](up);
\vertex [below=0.6cm of c](dw);
\vertex [left=0.6cm of up] (a);
\vertex [right=0.6cm of up] (b);
\vertex [left=0.6cm of dw] (a');
\vertex [right=0.6cm of dw] (b');
\diagram* {
(a) -- [plain] (b),
(a') -- [plain] (b'),
(up) -- [plain] (dw)
}; 
\end{feynman}
\end{tikzpicture}
}

\newcommand{\deltazerosmall}{
\begin{tikzpicture}[baseline]
\begin{feynman}
\vertex (c);
\vertex [above=0.3cm of c](up);
\vertex [below=0.3cm of c](dw);
\vertex [left=0.3cm of up] (a);
\vertex [right=0.3cm of up] (b);
\vertex [left=0.3cm of dw] (a');
\vertex [right=0.3cm of dw] (b');
\diagram* {
(a) -- [plain] (b),
(a') -- [plain] (b'),
(up) -- [plain] (dw)
}; 
\end{feynman}
\end{tikzpicture}
}

\newcommand{\deltazerosq}{
\begin{tikzpicture}[baseline]
\begin{feynman}
\vertex (c);
\vertex [above=0.6cm of c](up);
\vertex [below=0.6cm of c](dw);
\vertex [left=0.6cm of up] (a);
\vertex [right=0.6cm of up] (c2);
\vertex [right=0.6cm of c2] (b);
\vertex [left=0.6cm of dw] (a');
\vertex [right=0.6cm of dw] (c3);
\vertex [right=0.6cm of c3] (b');
\diagram* {
(a) -- [plain] (b),
(a') -- [plain] (b'),
(up) -- [plain] (dw),
(c2) -- [plain] (c3)
}; 
\end{feynman}
\end{tikzpicture}
}

\newcommand{\deltazerocub}{
\begin{tikzpicture}[baseline]
\begin{feynman}
\vertex (c);
\vertex [above=0.6cm of c](up);
\vertex [below=0.6cm of c](dw);
\vertex [left=0.6cm of up] (a);
\vertex [right=0.6cm of up] (c2);
\vertex [right=0.6cm of c2] (c4);
\vertex [right=0.6cm of c4] (b);
\vertex [left=0.6cm of dw] (a');
\vertex [right=0.6cm of dw] (c3);
\vertex [right=0.6cm of c3] (c5);
\vertex [right=0.6cm of c5] (b');
\diagram* {
(a) -- [plain] (b),
(a') -- [plain] (b'),
(up) -- [plain] (dw),
(c2) -- [plain] (c3),
(c4) -- [plain] (c5)
}; 
\end{feynman}
\end{tikzpicture}
}

\newcommand{\deltatwo}{
\begin{tikzpicture}[baseline]
\begin{feynman}
\vertex (c);
\vertex [above=0.6cm of c](up);
\vertex [below=0.6cm of c](dw);
\vertex [left=0.6cm of up] (a);
\vertex [right=0.6cm of up] (c2);
\vertex [right=0.6cm of c2] (c4);
\vertex [right=0.6cm of c4] (b);
\vertex [left=0.6cm of dw] (a');
\vertex [right=0.6cm of dw] (c3);
\vertex [right=0.6cm of c3] (c5);
\vertex [right=0.6cm of c5] (b');
\vertex [right=1.2cm of c] (h1);
\diagram* {
(a) -- [plain] (b),
(a') -- [plain] (b'),
(up) -- [plain] (dw),
(c4) -- [plain] (c5),
(c) -- [plain] (h1)
}; 
\end{feynman}
\end{tikzpicture}
}

\newcommand{\deltatwosmall}{
\begin{tikzpicture}[baseline]
\begin{feynman}
\vertex (c);
\vertex [above=0.3cm of c](up);
\vertex [below=0.3cm of c](dw);
\vertex [left=0.3cm of up] (a);
\vertex [right=0.3cm of up] (c2);
\vertex [right=0.3cm of c2] (c4);
\vertex [right=0.3cm of c4] (b);
\vertex [left=0.3cm of dw] (a');
\vertex [right=0.3cm of dw] (c3);
\vertex [right=0.3cm of c3] (c5);
\vertex [right=0.3cm of c5] (b');
\vertex [right=0.6cm of c] (h1);
\diagram* {
(a) -- [plain] (b),
(a') -- [plain] (b'),
(up) -- [plain] (dw),
(c4) -- [plain] (c5),
(c) -- [plain] (h1)
}; 
\end{feynman}
\end{tikzpicture}
}

\hypersetup{colorlinks, citecolor=bluscuro, linkcolor=bluscuro, urlcolor=bluscuro} 
\definecolor{rossos}{cmyk}{0,1,1,0.55}
\definecolor{bluscuro}{rgb}{0.15, 0.2, .85}
\definecolor{bluchiaro}{cmyk}{1,.3,0.,0.1}

\usepackage{multirow}
\usepackage[hmarginratio=1:1,top=32mm,columnsep=20pt]{geometry}

\usepackage[font=footnotesize, labelfont=bf]{caption}

\numberwithin{equation}{section}

\newcommand{\fref}[1]{Fig.~\ref{#1}} 
\newcommand{\eref}[1]{Eq.~\eqref{#1}}

\newcommand{\aref}[1]{App.~\ref{#1}}
\newcommand{\sref}[1]{Sec.~\ref{#1}}
\newcommand{\cref}[1]{Chapter~\ref{#1}}

\newcommand{\nnl}{\nonumber \\}

\newcommand{\beq}{\begin{equation}} 
\newcommand{\eeq}{\end{equation}} 
\newcommand{\ba}{\begin{array}}  
\newcommand{\ea}{\end{array}} 
\newcommand{\bea}{\begin{eqnarray}}  
\newcommand{\eea}{\end{eqnarray} }  
\newcommand{\be}{\begin{eqnarray}}  
\newcommand{\ee}{\end{eqnarray} }  
\newcommand{\bal}{\begin{align}}
\newcommand{\eal}{\end{align}}   
\newcommand{\bi}{\begin{itemize}}  
\newcommand{\ei}{\end{itemize}}  
\newcommand{\ben}{\begin{enumerate}}  
\newcommand{\een}{\end{enumerate}}  
\newcommand{\bc}{\begin{center}}
\newcommand{\ec}{\end{center}} 
\newcommand{\bt}{\begin{table}}
\newcommand{\et}{\end{table}}  
\newcommand{\btb}{\begin{tabular}}
\newcommand{\etb}{\end{tabular}}

\newcommand{\cM}{{\mathcal M}}

\newcommand{\lambdabar}{{\mkern0.75mu\mathchar '26\mkern -9.75mu\lambda}}

\def\lpl{\, \lambda_{\rm Pl}}

\def\ra{\rangle}
\def\la{\langle}  

\newcommand{\dint}[2]{\frac{d^{#1} #2}{(2 \pi)^{#1}}}
\def\dd{\delta\!\!\!{}^-\!}

 \def\mpl{m_\mathrm{Pl}}
 \def\lambdapl{\lambda_\mathrm{Pl}}

\begin{document}
\begin{center}

\vspace*{-25mm}

\begin{flushright}
{\small Saclay-t22/086}\\
\end{flushright}

\vspace{2cm}
{\Large \bf Classical vs Quantum Eikonal Scattering
and its Causal Structure}
\vspace{1.4cm}\\
{Brando Bellazzini$\,^{1}$, Giulia Isabella$\,^{1,2,3}$, 
Massimiliano Maria Riva$\,^{1, 4}$}

 \vspace*{.5cm} 
\begin{footnotesize}
\begin{it}
 $^1$ Universit\'e Paris-Saclay, CNRS, CEA, Institut de Physique Th\'eorique, 91191, Gif-sur-Yvette, France.  \\
 $^2$ Universit\'e Paris-Saclay, CNRS/IN2P3, IJCLab, 91405 Orsay, France\\
   $^3$ D\'epartment de Physique Th\'eorique,
Universit\'e de Gen\`eve, CH-1211 Gen\`eve, Switzerland\\
  $^4$ Deutsches Elektronen-Synchrotron DESY, Notkestrasse 85, 22607 Hamburg, Germany\\
\end{it}
\end{footnotesize}

\vspace*{.2cm} 

\vspace*{10mm} 
\begin{abstract}\noindent\normalsize
We study the eikonal scattering of two gravitationally interacting bodies, in the regime of large angular momentum and large center of mass energy. 
We show that eikonal exponentiation of the scattering phase matrix is a direct consequence of the group contraction $SU(2)\to ISO(2)$, from rotations to the isometries of the plane,  in the large angular momentum limit. We extend it to all orders in the scattering angle, and for all masses and spins. 
The emergence of the classical limit is understood in terms of the continuous-spin representations admitted by $ISO(2)$.  
We further investigate the competing classical vs quantum corrections to the leading classical eikonal scattering, and find several interesting examples where quantum corrections are more important than Post-Minkowskian's.  As a case of study, we analyse the scattering of a photon off a massless neutral scalar field, up to next-to-leading order in the Newton constant, and to leading order in the fine structure constant. We investigate the causal structure of the eikonal regime and establish an infinite set of non-linear positivity bounds, of which positivity of time delay is the simplest.
\end{abstract}
\end{center}
\newpage
{ \hypersetup{hidelinks} \tableofcontents }

\newpage

\section{Introduction}
In this paper we study  the large-energy and large-angular momentum limit of the two body scattering problem, known as eikonal scattering. We investigate in particular transplanckian gravitational scattering at large impact parameter, although the scope of our study  is wider and it applies as well beyond gravitational interactions. 

We are interested in both the classical and quantum aspects, and we are going to contrast their features.  Classical eikonal effects control the leading contributions to the observables and are of direct interest for e.g. astrophysical bodies, while we show that quantum (non-gravity) effects may become the dominant corrections whenever particles of the Standard Model (SM) are scattered at transplanckian energy and large impact parameter.  

The goal of the present work is twofold:  {\it i)} providing a systematic Effective Field Theory (EFT) expansion in the eikonal regime that works to any desired order in the ratios of length scales, for particles of any mass and spin, and large gravitational coupling;  and 
{\it ii)} understanding the causal structure in a theory with dynamical gravity, including nevertheless the leading ---resolvable--- quantum fluctuations from the dynamics of particles.   It turns out that these quantum field theory (QFT) corrections are orders of magnitude more important than semiclassical corrections for particles or small black holes scattering.  

 The eikonal approximation has a long history that started in non-relativistic quantum mechanics inspired by the geometric optics limit, see e.g. \cite{landau3}.    One of most interesting applications in modern times was the study of transplanckian scattering  of massless particles in a beautiful series of works \cite{Amati:1987uf,Amati:1987wq,Amati:1990xe,Amati:1992zb,tHooft:1987vrq,Verlinde:1991iu,Kabat:1992tb}, where one of the  original motivations was to construct explicitly an unitary $S$-matrix to address the black hole information paradox.\footnote{Transplanckian eikonal scattering has found applications even within particle phenomenology at colliders, see \cite{Giudice:2001ce}.}  Despite the original goal was  beyond reach, the eikonal has recently found a new life with the detection of gravitational waves (GWs) that has sparked a fervent activity in the application of such method to study the scattering of massive compact objects such as black holes or neutron stars, see e.g. Refs. \cite{Bern:2019crd,Kalin:2020mvi,Mogull:2020sak,Cheung:2020sdj,Herrmann:2021tct,DiVecchia:2021bdo,Brandhuber:2021eyq,Bjerrum-Bohr:2021vuf,Buonanno:2022pgc,Bastianelli:2021nbs}, and references therein for an incomplete list of works on this subject. 

Since gravity couples to the energy-momentum tensor, the gravitational coupling is very large in the transplanckian regime, in fact tremendously large for astrophysical bodies. 
Yet the theory remains under theoretical control because  the pull of gravity decays with distance in $D\geq 4$ spacetime dimensions. 
Therefore one can expect that an EFT expansion in the relevant length scales ---Schwarzschild radius $R_s$, body sizes $L_\odot$,  Compton wavelengths $\lambdabar$ etc--- over the impact parameter $b$ would allow to control and organize the infinitely many Feynman diagrams required with such a large gravitational coupling. 

In this paper we show that such an EFT expansion at large center of mass (c.o.m.) energy ---hence large coupling--- and large impact parameter, actually corresponds to study systematically the corrections to the limit $\ell\to\infty$ of the partial wave decomposition of scattering amplitudes for particles of any mass and spin. In particular, we show that the eikonal (matrix) exponentiation of the scattering phase, that is obtained by resumming infinitely many diagrams  as depicted in \eqref{eikonalintro}, 
\begin{equation}
\begin{aligned}
\label{eikonalintro}
e^{2i \delta(s,\mathbf{b})} = \mathrm{Exp}\left\{\raisebox{4pt}{\deltazerosmall}+\raisebox{4pt}{\deltatwosmall}+ \, \dots \right\} = \mathbb{I} \!+\!\!\! \scalebox{0.7}{\raisebox{3pt}{\deltazero}} \!\!+\!\! \scalebox{0.7}{\raisebox{3pt}{\deltazerosq}}  \!\!+\!\! \scalebox{0.7}{\raisebox{3pt}{\deltazerocub}} \!\!\!+ \dots +\!\!\! \scalebox{0.7}{\raisebox{3pt}{\deltatwo}} +\dots 
\end{aligned}
\end{equation}
can be understood as a direct consequence of the  $SU(2)\to ISO(2)$ group contraction, where the isometries of the sphere (3D rotations) are well approximated by those of a tangent plane at large radius (euclidean 2D translations and rotations). 
In this sense, we show that eikonal expansion holds for the same reason a flat-earth model of a our planet is an excellent approximation over length scales much smaller than the earth radius.\footnote{To the point  flat-earth believers  extrapolate that approximation beyond its range of validity.} The continuous-spin irreducible representations (irreps) of the non-compact group $ISO(2)$ corresponds precisely to the emergence  of a continuous ---classical--- angular momentum in the eikonal limit. 
Although we derive an all-order eikonal expansion in \sref{Sec:allorders},  we do not know what the radius of convergence may be, the optimistic expectation based on black hole formation being $R_s/b\sim 1$. 

While the leading transplanckian eikonal scattering is semiclassical, we show that there is a huge window in length scales where the quantum corrections $\sim \alpha (\lambdabar/b)^2$ controlled by some QFT coupling $\alpha$ and Compton wavelength $\lambdabar$ (other than irrelevant quantum gravity $(\lpl/b)^2$) are much more important than semiclassical post-Minkowskian $(R_s/b)^2$-corrections, as soon as $ \sqrt{\alpha} \lambdabar> R_s$. Likewise, if very light, essentially massless, particles run in loops  their relative corrections  $\sim \alpha \log^2 b/\lambdabar$  may well dominate the semiclassical ones $(R_s/b)^n$ whenever $\alpha\gg (R_s/b)^2$.  For instance, when scattering SM particles at transplanckian energy, there are about fifteen orders of magnitudes where SM quantum loops can be more important than post-Minkowskian corrections.  This is usually obscured in the literature by conflating the eikonal scattering with the blind classical limit $\hbar\to 0$ limit, despite $\alpha \lambdabar$ may in fact be larger than $R_s$.\footnote{Needless to say,  this is unlikely to be the case  for scattering astrophysical bodies unless considering modified-gravity theories with e.g. an extra light particle coupled to matter and with Compton wavelength larger than the $R_s$  or $L_\odot$. }

The fact that quantum effects may represent the leading corrections in transplanckian scattering opens  the  possibility to study the role of quantum fluctuations in the context of dynamical gravity in the controlled setting of the eikonal expansion. 
We investigate in particular the causal structure of scattering amplitudes in the eikonal regime including QFT corrections, significantly extending the results of \cite{Camanho:2014apa} along the lines of \cite{Bellazzini:2021shn, Bellazzini:2020cot,Caron-Huot:2021rmr,Caron-Huot:2022ugt}.  

In particular, focusing on the instructive (and phenomenologically interesting) graviton-scalar scattering,  we prove an infinite tower of non-linear positivity constraints on the EFT, assuming analyticity and unitarity of the scattering amplitudes.   These constraints represent the imprint of causality on infrared observables. Notice that we are neglecting spooky quantum-gravity corrections $\sim(\lpl/b)^2$, so that analyticity of amplitudes is a rather conservative assumption.  
Positivity of the time delay, that is ``asymptotic causality'',  is just the simplest of these positivity bounds.  
Higher energy derivatives of the scattering phase are constrained as well but, in the eikonal limit, the bounds requires also taking the rigid limit of fixed and weak gravitational background. 


This work is organised as follows. Using general and simple scaling arguments, in \sref{sec:generalities} we introduce an operative way to study the eikonal regime based on the large angular momentum limit. We propose also a better way of  contrasting classical vs quantum effects, based on their size and resolvability with respect to physical observables such as the deflection angle. \sref{Sec:Eikonal} is devoted to prove and extend the eikonal exponentiation \eqref{eikonalintro} via partial waves decomposition for particles of any mass and spin.  In particular, in \sref{Sec:allorders} we show how to extract an eikonal expansion valid to all orders in the deflection angle. Then we discuss an application of this in \sref{Sec:transplanckian} by studying  the scattering of a photon against a massless neutral scalar, up to next-to-leading order in $G$ (Newton constant)  and at the leading order in the fine structure constant $\alpha$. In \sref{section:CausalityPositivity} we study causality constraints in the form of positivity bounds for the EFT of scalar-graviton scattering. We present our conclusions in \sref{sec:concl}.  Certain details on infrared divergences,  on the partial wave expansion, and the 2-loop calculations can be found in appendices \ref{wordline-Eik}, \ref{AppPhases}, and \ref{app:MIs}  respectively.

\section{Eikonal and Transplanckian Scattering}
\label{sec:generalities}

We consider the gravitational scattering of two ---possibly spinning--- bodies of incoming momenta $p_1$ and $p_2$  and denote by $s=(p_1+p_2)^2$ and $t=q^2$ the Mandelstam variables for the c.o.m. energy and the momentum exchanged, squared.\footnote{In order to simplify the kinematics we consider the case where the final masses are $m_3=m_1$ and $m_4=m_2$, although the bodies may  change flavor, spin, etc. in the scattering. We call $u=2m_1^2+2m_2^2-s-t$ the other Mandelstam variable.  }

There are four  length scales in this problem. Two are kinematical ---the Compton wavelength $\lambda_s \equiv 1/\sqrt{s}$ and the impact parameter $b$ (conjugate variable to $q$, more on this in the next sections)--- whereas the other two are dynamical:  the Planck length $\lpl$ and  the effective  Schwarzschild radius $R_s$, namely\footnote{Reintroducing $\hbar$ while keeping $c=1$,  the planck length is $\lpl=\sqrt{G\hbar}$ and $\mpl^2=\hbar^2/(8\pi\lpl^2)$.  }
\begin{equation} 
		R_s \equiv 2 G \sqrt{s} \, ,\qquad 
		\lpl=\sqrt{G} =\frac{1}{\mpl\sqrt{8\pi}}\, .
		\label{eq:Rs_def}
\end{equation}
The $\mpl$ is the reduced Planck mass and $G$ the Newton constant. 

The $R_s$ and $\lpl$ mark the onset  of classical strong non-linearities (e.g. black hole formation) and strong quantum-gravity (QG) effects, respectively.    
The impact parameter measures the orbital angular momentum $b|\bm{p}|$,  where $\bm{p}$ is the c.o.m. 3-momentum.\footnote{It can be expressed in terms of Mandelstam invariants as $\bm{p}^{2}=[s-(m_1+m_2)^2][s-(m_1-m_2)^2]/4s$.}  In the following we consider bodies with spin projections  much smaller than the orbital spin, so that the impact parameter can be traded for the total angular momentum $\ell\simeq b|\bm{p}|$, which is taken large. 

Other dynamical length scales, collectively denoted by $\lambdabar$ hereafter, may appear in the problem representing e.g. the size of the  (astrophysical) body, the compositeness-scale of particles, the Compton wavelength of virtual particles, the string length, etc.  

The problem is further caractherized by the strength $\alpha_g$ of the gravitational interactions
 \begin{equation}
 \label{alphaGdef}
 \alpha_g \equiv R_s  \frac{(p_1\cdot p_2)^2}{s|\bm{p}|} \sim R_s \sqrt{s} \propto Gs\, 
 \end{equation}
defined proportionally to the residue of the elastic amplitude at $t=0$, such that the scattering phase at large angular momentum is $\delta_\ell= \alpha_g \log\ell$ for $\ell\to\infty$. The right-most part of \eqref{alphaGdef} shows the scaling of $\alpha_g$ for massless particles. For future reference, in the probe limit the scaling is instead $\alpha_g=R_s|\bm{p}|$.  
The gravitational coupling controls two regimes: 
 \begin{equation} 
 \begin{array}{cccc}
\mbox{\it subplanckian } &  s\ll \mpl^2 \Longrightarrow     &  \alpha_g\ll 1 &    R_s \ll \lpl \ll \lambda_s \\
&&&\\
\mbox{\it transplanckian } &  s\gg \mpl^2 \Longrightarrow &  \alpha_g \gg 1 & \lambda_s\ll \lpl \ll  R_s\, .
\end{array}
 \end{equation}
 
In subplanckian  scattering, the bodies need full fledge quantum description because the Compton wavelength $\lambda_s$ is larger than the classical $R_s$;  yet the gravitational interactions are still very weak, so that they can be systematically taken into account by deforming the free (quantum) theory, e.g. in the Born approximation scheme.  If other interactions are also weak, textbook QFT pertubative calculations in terms of a finite numbe of Feynman's diagrams are enough to achieve any desired accuracy. 

Transplanckian scattering ---reached for extremely boosted particles relative to one another or just for very massive (e.g. astrophysical) bodies---  is characterised instead  by a large coupling $\alpha_g\gg 1$, and the perturbative approximation scheme with a  finite number of Feynman's diagrams is no longer viable.  

However, in transplanckian scattering,  one can adopt an EFT expansion thanks to the large hierarchy of length scales 
 \begin{equation}
\frac{ \lambda_s}{b}  \ll  \frac{\lpl}{b}  \ll  \frac{R_s}{b} \ll 1
 \end{equation}
 as long as the impact parameter $b$ is taken much larger than the Schwarzschild radius.  
  In other words, there is another expansion parameter, $R_s/b\ll 1$, which is nothing but the classical scattering angle itself~\footnote{Explicitly,  $\theta\simeq q/p\sim F b/p\sim \alpha_g/pb\sim R_s/b$ where $F$ is the force.}  once expressed in terms of the conjugate variable $q$, namely \begin{equation}
	\theta \simeq \frac{|\bm{q}|}{|\bm{p}|} \sim \frac{R_s}{b } \ll 1 \, . 
	\label{eq:smalldef}
\end{equation}
The exchanged momentum must therefore be very small to allow the EFT expansion to work in the large coupling regime.   In a sense, the two bodies have a tremendous center of mass energy ---a large gravitational  coupling--- but the pull of gravity decays with distance so that a large transverse separation (large angular momentum) allows us to control the scattering of nearly deflected bodies.  As long as $\ell$ is taken large and the angle is small, one can control the theory in all energy range, from the subplanckian (small coupling) to  the transplanckian (large coupling) regime. Equivalently, the eikonal regime corresponds to the kinetic energy being much larger than the potential energy,  $\sqrt{s}\gg Gs/b$, which immediately implies \eqref{eq:smalldef}.

In the next section we address this eikonal regime of large impact parameter and small angle by systematically studying the limit
\begin{equation}
  \label{LimitContraction2}
\ell\rightarrow \infty\,,\qquad  \theta\rightarrow 0\,, \qquad \ell\theta\rightarrow \mathrm{fixed}\sim \alpha_g 
 \end{equation}
 in the partial wave decomposition, and its corrections at finite $\ell$ and $\theta$. The leading effect corresponds to resum an infinite class of Feynman diagrams (ladder and cross-ladder), whereas the next-to-leading contributions keep track of the $O(R_s/b)^n$- corrections and, possibly, of other small-coupling effects.   In \sref{Sec:allorders} we provide the all-order eikonal amplitude which can be used to extract any desired $O(R_s/b)^n$, extending thus the result to finite scattering angle. \\

Since $R_s$ controls the classical non-linear gravitational effects, and because it is much larger than the quantum gravity scale $\lpl$ and the Compton wavelength $\lambda_s$,  the transplanckian regime is usually conflated with the semi-classical limit. 

This identification of transplanckian scattering with classical physics is, however, too quick and sometimes misleading. Only the leading term needs to be classical. 
There could be indeed other length scales in the problem $\lambdabar$, e.g. other particles Compton wavelengths,  which are not necessarily small relatively to $R_s$ or even to $b$ as one lowers it:  
\begin{equation}
\lambdabar \ll R_s \ll b \quad  \mbox{vs} \quad  R_s \ll \lambdabar \ll b  \quad \mbox{vs} \quad    R_s \ll  b \ll  \lambdabar   \, .
\end{equation}
Only the leftmost inequalities allow to consistently neglect  $\lambdabar/b$ corrections relative to $R_s/b$. 

\begin{figure}[t]
\centering
\includegraphics[width=0.8\linewidth]{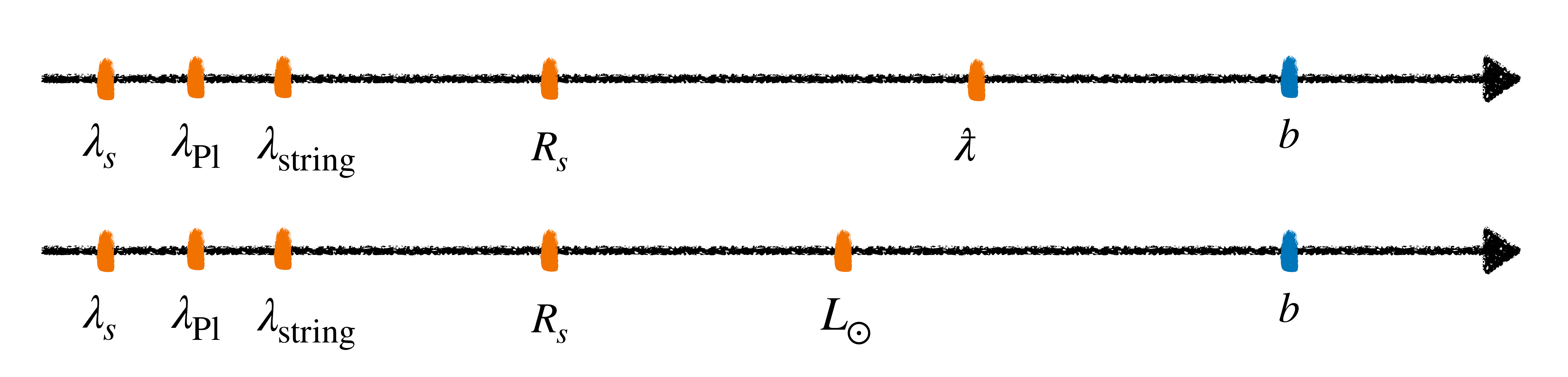}
\caption{{\small Scale lengths of the system. Bottom: tidal relative corrections $(L_\odot/b)^n$ are largest and dominate the modifications to the leading eikonal. Top: the most important corrections to leading eikonal arise from higher-derivative operators generated by particles of Compton wavelength $\lambdabar=1/m_e$ that are running in loops, as long as $\alpha \lambdabar^2\gg R_s$ and $b\gg \lambdabar$. For $b\ll \lambdabar$ resummation to all orders in $\lambdabar/b$ is needed, corresponding to work with a new EFT where new degress of freedom are propagating. A typical example of $\delta\theta/\theta$, to first order in the coupling constant,  is $\sim \alpha \log^2 \lambdabar/b$ in this regime. }}
\label{fig:scales}
\end{figure}

 And there can be quantum corrections due to other weak couplings, collectively denoted hereafter by $\alpha$ (e.g. a gauge coupling in  a QED-like theory), which can be larger than the classical corrections $O(R_s/b)^n$ to the leading term. 
This never happens for astrophysical bodies (unless there were e.g. new light degrees of freedom with Compton wavelength larger than $R_s$), but it can definitely happen for scattering {\it particles} at transplanckian energies, or even for black holes of radius much larger than $\lpl$ but smaller than the various Compton's wavelengths of particles in the SM.  See Figure~\ref{fig:scales} for an illustration. 

Consider for example a theory of photons coupled to gravity,  integrating out a charged particle of spin $s_e$ and mass $m_e$ at one-loop the theory is matched to an extra effective operator  (among others) 
\begin{equation}
\label{eqExampleFFRie}
\delta\mathcal{L}= c \left(\frac{\alpha}{4\pi} \lambdabar^2\right) F_{\mu\nu}F_{\rho\sigma}R^{\mu\nu\rho\sigma} 
\end{equation}  
where $\lambdabar=1/m_e$ and $c=\pm( 2s_e+1)/180$ with  $+$ ($-$) for a boson (fermion).  In the regime $ R_s \ll \sqrt{|c|\alpha/4\pi}\lambdabar \ll b$, the operator \eqref{eqExampleFFRie}  changes the scattering angle by a relative amount $O(c\alpha/4\pi\, \lambdabar^2/b^2)$ which is more important than the $O(R_s/b)^2$ classical corrections \cite{Bellazzini:2021shn}. 
The regime $R_s<b<\lambdabar$, which is beyond the EFT description \eqref{eqExampleFFRie}, is also discussed in \cite{Bellazzini:2021shn} using full QED coupled to gravity, and it is further extended  in \sref{Sec:transplanckian}.

For other  examples where $\lambdabar=\lambda_{\mathrm{string}}$ in string theory,  see \cite{Amati:1987wq,Amati:1987uf}  for the regime  $R_s \ll \lambda_{\mathrm{string}} \ll b$, the \cite{Amati:1990xe} for $\lambda_{\mathrm{string}} \ll R_s \ll b$, and \cite{Amati:1988tn,Amati:1987uf} for $R_s<b<\lambda_{\mathrm{string}}$.  We collectively refer to this bulk of papers as the ACV reference.  
 For more recent results see also appendix~E of Ref.~\cite{Camanho:2014apa} and \cite{DAppollonio:2015fly},  again in the non-trivial regime $R_s<b<\lambda_{\mathrm{string}}$. 
 Needless to say, to address the $b\ll \lambdabar$ case,  one needs to incorporate new propagating degrees of freedom that appear from $b\simeq \lambdabar$ downward.
 
 We remark that it is possible to extend the eikonal regime below $\lambdabar$ for strings and particles because the degrees of freedom running in loops  carry at most momenta of planckian order (past which the integrals converge), which is still far from being enough to compete with the orbital contribution, i.e. $b |\bm{p}|\gg \mpl \lambdabar$, for $s\gg\mpl^2$. This is generically not the case instead for astrophysical bodies where the constituents that become visible  at $\lambdabar=L_{\odot}$ are themselves transplanckian and carry an O(1) fraction of the total angular momentum; in this case the eikonal approximation breaks down. 
 \\

 As we explain in detail in \sref{sec:classicalT}, transplanckian scattering is indeed semiclassical but in the sense that $\theta$ and the angular momentum $\ell\sim b|\bm{p}|$  have simultaneously small quantum uncertainties.  Semi-classicality does not mean that relative corrections to the leading value of $\theta$ must be classical in origin too. They can well be dominated in fact by quantum loops $O(\alpha/4\pi)$, rather than by classical effects $O(R_s/b)$, like in the example \eqref{eqExampleFFRie}.  Likewise, corrections due to other length scales $\lambdabar/b$ could also dominate over the classical corrections $O(R_s/b)$. 
 
   Reintroducing momentarily $\hbar$, the customary  approach to select the relevant scales in the two-body problem is looking at the $\hbar$-scaling of scattering amplitudes, which assigns what is ``classical" and what is ``quantum". 
   We think  that a more appropriate terminology to take into account a larger number of scales is actually {\it ``resolvable'' }  and {\it ``non-resolvable''} effects. Indeed, we argue in \sref{sec:classicalT} that certain quantum effects such as those produced by couplings with the SM are in fact resolvable and possibly more important than certain classical contributions. 
      
   Alternatively, whenever taking $\hbar\to 0$ limit one should determine what to keep fixed depending on the relative size of the various contributions.  Sending to zero the Compton wavelength $\lambdabar=\hbar/m_e$ of particles running in a loop,   or the (dimensionaless) loop expansion parameter itself $\alpha/4\pi=g^2\hbar/16\pi^2$, does not really make sense in transplanckian scattering whenever $\lambdabar >R_s$ or $\alpha/4\pi> (R_s/b)^2$.
   
   We insist on the fact that the relative hierarchy between those scales completely depends on the problem studied. A system involving the sun, for instance,  is dominated by finite size $\lambdabar=L_\odot$ effects, as its radius $L_\odot$ is much larger than its Schwarzschild radius and any other $\lambdabar$. In this case the classical limit emerges as the leading effect in the $\hbar\to 0$ scaling, see e.g. \cite{Kosower:2018adc}.   On the other hand, for small black holes, say $\lpl \ll R_s \ll \mathrm{1/TeV}$, the relevant corrections would actually come from SM couplings where the  SM particles have $\lambdabar> R_s$. Likewise, SM loops are the most important corrections for  transplanckian scattering of photons at small impact parameter (but still larger than $R_s$).  \\

We would like to mention that the eikonal regime exists and it is relevant for instance also in QED without gravity.  In the scattering of two particles of charge $Z_i$, the condition of kinetic energy to dominate over the potential energy  leads to $\omega \ll \alpha Z_1 Z_2/b$ and, analogously to the gravitational case, the eikonal regime in QED is also characterized by large total angular momentum taking the form $\ell \sim \alpha_{\rm em}Z_1 Z_2/\theta$. Contrary to the gravitational case, however, the classical limit $\ell\theta\gg 1$ is reached only for very large $Z_i$, corresponding necessarily to non-elementary objects such as heavy nuclei\footnote{The calculation of the Lamb shift due to nuclei with $Z_i\gg 1$, see e.g.\cite{Weinberg:1995mt},  is the QED bounded-orbit analog of the quantum corrections to the gravitational unbounded orbits we calculate in the next sections. }, charged black holes, etc.

 \subsection{Classical vs Quantum and Resolvability}
 \label{sec:classicalT}
 
As we have discussed so far, the eikonal scattering may apply to both subplanckian and transplanckian regimes, as long as the angular momentum is large and the scattering angle is small.  It is however only for transplackian energies that  eikonal scattering is also semi-classical as we show in this subsection, along with the classification of the next-to-leading classical and quantum corrections. 

We can talk of semi-classical trajectory whenever we can simultaneously assign the impact parameter  $b$ and the resulting deflection angle $\theta$, i.e. if the outcome of the scattering experiment  is approximately deterministic because of small quantum uncertainties: 
  \begin{equation}
  \Delta\theta/\theta\ll 1 \qquad \mathrm{\&}\qquad \Delta b/b\ll 1\,\, .
  \label{eq:Uncert}
  \end{equation}
The quantum uncertainties can be related as  
\begin{equation}
\label{quantumindet1}
\frac{\Delta \theta}{\theta}\sim \frac{\Delta q}{ \bm{p} \theta} \gtrsim \frac{1}{\Delta b} \frac{1}{\boldsymbol{p} \theta} \sim \frac{1}{\Delta b/b} \frac{1}{ \ell \theta}  
\end{equation}
so the semiclassical limit  requires $\ell\theta\gg 1$.  Moreover, in the eikonal  scaling discussed in the previous subsection,  $\ell \theta\sim \alpha_g$ and therefore classicality requires tranplanckian scattering $\alpha_g \gg 1$ (that is $\delta_\ell\gg1$) namely 
\begin{equation}
	\frac{\Delta \theta}{\theta}\frac{\Delta b}{b} \sim \frac{1}{\alpha_g} \ll 1 \, .
\end{equation}
 The intersection between  semiclassical limit and eikonal scattering allows us to describe transplanckian eikonal scattering  in terms of trajectories in spacetime. Notice also that being transplanckian implies 
\begin{equation}
	\frac{1}{G s } \sim \frac{\lambda_s}{R_s} \ll 1 \, ,
\end{equation}
which is equivalent to 
\begin{equation}
	\frac{\lpl^2}{R_s^2} \ll 1 \, .
\end{equation}
In classical transplanckian scattering, $R_s$ and $\lpl$ are hierarchical hence the latter matters the least in the EFT. \\

Let us now take the leading order deflection angle $\theta \sim R_s/b$ and imagine  including  possible subleading corrections $\delta \theta$. It is relevant at this point to ask which corrections are {\it resolvable}: whether their effect  is larger than quantum uncertainty. 

We consider the following type of modifications: $\delta \theta \sim\left(R_s/b\right)^n$, $\delta \theta \sim\left(\lpl/b\right)^{2n}$ and $\delta \theta \sim \alpha^n (\lambdabar/b)^k$, where we recall that $\alpha$ represents some other coupling in the theory, e.g. $\alpha=\alpha_{\mathrm{em}}$ the fine structure constant of QED.  Tidal corrections $(L_\odot/b)^k$ fit in the  $\alpha^n (\lambdabar/b)^k$-classification  with  $\lambdabar=L_\odot$ and $\alpha$ some analog of e.g. the love number.  

The first type $\left(R_s/b\right)^n$ are classical GR corrections, sometimes referred to as Post-Minkowskian (PM) corrections. The second one $\left(\lpl/b\right)^{2n}$ are purely quantum gravity (QG) corrections. The last type of modifications  $\alpha^n (\lambdabar/b)^k$ may be ``quantum'' corrections whenever controlled by a coupling such as $\alpha=\alpha_{\mathrm {em}}$ (for this we dub them ``gauge'', but the coupling  can certainly be a Yukawa squared or else), but they can  also be classical  whenever due to finite-size ---tidal--- corrections.  To summarise\footnote{We are implicily assuming that there are no other long range interactions among the two bodies, it is simple to amend this simplifying assumption.}   
\begin{equation}
	\frac{\delta \theta}{\theta} \sim 
	\begin{cases}
		\left(R_s/b\right)^n & \text{PM,} \\
		\left(\lpl/b\right)^{2n} & \text{QG,} \\
		\alpha^n \left({\lambdabar}/{b}\right)^k  & \text{Gauge/Tidal} 
	\end{cases}
\end{equation}
where $k=0$ for $n>0$ corresponds to powers of $\log b/\lambdabar$ factors (typically arising from light particles loops).  From  \eqref{quantumindet1}, we get
\begin{equation}
\frac{\Delta \theta/\theta}{\delta \theta/\theta} \gtrsim \frac{1}{\Delta b} \frac{1}{\boldsymbol{p}\, \delta\theta} \sim  \frac{1}{\Delta b/b} \frac{1}{\ell\theta}\left(\frac{1}{\delta\theta/\theta}\right)\sim
\begin{cases}
 \frac{1}{\Delta b/b} \frac{1}{\alpha_g\, \left(\frac{R_s}{b} \right)^{n}} & 
 \text{PM,}\\
  \frac{1}{\Delta b/b} \frac{1}{ \left(\frac{R_s}{b} \right)^2 \left(\frac{\lambda_{Pl}}{b} \right)^{2n-2}} & \text{QG,}\\
   \frac{1}{\Delta b/b} \frac{1}{ \alpha_g \alpha^n \left(\frac{\lambdabar}{b}\right)^k} & \text{Gauge/Tidal}
    \end{cases}  
    \label{eq:scalingQuantum}
\end{equation}
where we used $\ell\theta\sim \alpha_g$ and $(\lpl^2/b)^2 \alpha_g \sim (R_s/b)^2$. 

Let us comment on the different scenarios. For $\alpha_g$ large enough, the PM higher-order corrections lead to  resolvable effects\footnote{An amusing observation  is the fact that given explicit kinematic parameters, there exists a maximum PM-order $\bar{n} \sim \log\alpha_g /\log(b/R_s)$ which makes sense to calculate, beyond which quantum uncertainty takes over. }
\begin{equation}
\frac{\Delta \theta/\theta}{\delta \theta/\theta}  \ll 1 \, .
\end{equation} 
These are the corrections considered for instance in \cite{Bern:2019crd,Kalin:2020mvi,Mogull:2020sak,Cheung:2020sdj,Herrmann:2021tct,DiVecchia:2021bdo,Brandhuber:2021eyq,Bjerrum-Bohr:2021vuf}.

On the other hand, QG corrections  controlled by $\lpl/b$, the second equation in \eqref{eq:scalingQuantum}, are always smaller than the quantum uncertainty, as the eikonal limit requires $R_s/b\ll 1$, while being  transplanckian (semi-classical) imposes that $\lambda_{Pl}/b$ is even smaller. Therefore, one important lesson is that in eikonal scattering 
$$\mbox{{\it Quantum gravity corrections are never resolvable.}}$$ 
This implies  that $\delta\theta/\theta=O(\lpl/b)^{2n}$ corrections for scattering e.g. massless scalars in \cite{Amati:1990xe}, or the  quantum gravity corrections to bending of light \cite{Bjerrum-Bohr:2014zsa,Bai:2016ivl}, do not actually correspond to any physical effect such as  e.g. the alleged violations of the equivalence principle.   They just fall behind the intrinsic quantum fuzziness wall. 
Those corrections, however, are needed to correctly extract the resolvable 3PM corrections, as originally stressed and done in \cite{Amati:1990xe}. Of course,  mixed corrections scaling as $(\lambda_{Pl}/b)^n(R_s/b)^k$ are also not resolvable.

 This does not mean, however, that all quantum effects are necessarily negligible relative to the classical ones. For instance, from the third line of \eref{eq:scalingQuantum} we understand that there exists again a large enough $\alpha_g$ such that quantum effects due to gauge interactions $\delta\theta/\theta=\alpha^n (\lambdabar/b)^k$ are in fact resolvable in the eikonal-transplanckian limit.   In particular there is a regime in which such effects are even larger than the PM ones, see e.g. the effect of \eqref{eqExampleFFRie} in the regime $R_s\ll \sqrt{|c|\alpha/4\pi}/m_e$   \cite{Bellazzini:2021shn}. 
 We discuss and extend  this example in \sref{Sec:transplanckian}. 
 
 Even in the QED setting, it may happen than quantum corrections are generically more important than classical tidal corrections, as the Lamb shift contribution to the energy level in a bound state can easily be larger than the classical tidal effects, parametrically scaling as $\left(a_{\mathrm{nucl}}/a_{\mathrm{Bohr}}\right)^2$, due to finite nucleus size $a_{\mathrm{nucl}}$ with respect to the Bohr radius  $a_{\mathrm{Bohr}}$. 
 
 In fact, even tidal corrections that one would ascribe to classical physics may well be quantum in origin, as e.g. the size of a neutron star is controlled by its fully quantum equation of state, not to mention the stability of matter, atoms and nuclei.  Quantum vs classical is actually a false dichotomy: what really matters whenever dealing with the dimensionful parameter $\hbar$ are ---of course--- dimensionless ratios that can be large or small, and whether the effects are resolvable or not.

  We end this section with the second lesson: 
 $$\mbox{{\it Quantum (non-gravity) corrections may be resolvable and even leading over} PM.}$$

\section{The Eikonal Scattering as a Contraction}
\label{Sec:Eikonal}

In this section we show how the eikonal \eqref{eikonalintro} is recovered from the large angular momentum limit of the partial waves decomposition, for all masses and spins. 

We first review the main properties of the amplitude expansion on partial waves for spinning particles in \sref{sec:summarypartial}. Then, applying the limit \eqref{LimitContraction2}, we recover \eqref{eikonalintro}, which is physically understood as the  (c.o.m. little-)group contraction from $SU(2)$ to $ISO(2)$.  

 The continuous representations of $ISO(2)$ turn out to be crucial to recover the integral form of \eqref{eikonalintro}. This geometrical point of view of the eikonal is detailed in \sref{contractionSection}. Next, we focus on the subleading corrections to the eikonal limit, and obtain an all order expression \eqref{allordersQ} and its inverse \eqref{EikonalAllOrders}, that allows to extend the results to finite $\theta$. We conclude the section by extracting deflection angle and the time delay from the  eikonal amplitude for particles of any spin. 

\subsection{Partial Waves Reminder}
\label{sec:summarypartial}

From the classical arguments and scaling discussed in \sref{sec:generalities}, we expect to recover the eikonal exponentiation  \eqref{eikonalintro} in the limit of large total angular momentum. It is then natural to work in a basis of definite angular momentum, that is projecting the amplitudes on partial waves. As this partial wave decomposition plays an important role in the following, we review its salient aspects in this section, leaving a detailed derivation to  Appendix~\ref{AppPhases}.

Firstly, we build two-particle states by taking the tensor product of irreps of Poincar\'{e} in the c.o.m. frame.  For future convenience, we choose a net angle $\theta$ with the $z$-axis equally split between the incoming and outgoing state. The azimuthal angle $\phi$ takes into account rotations with respect to the $xz$-plane.
As usual, the states are labeled by momentum $p_i$, spin $S_i$, helicity $\lambda_i$ (spin projection along $\bm{p}_i$)  and other quantum numbers, collectively denominated $\alpha_i$
 \begin{align}
 \label{statesdef1}
& | 1^{\lambda_1} \,  2^{\lambda_2} \rangle  \equiv
R(\phi,-\frac{\theta}{2},-\phi) | p_1\, S_1\, \lambda_1, \alpha_1 ; p_2\, S_2\, \lambda_2, \alpha_2 \rangle_{\substack{ \\ c.o.m. \\  \!\theta=\phi=0}}\,, \\ 
 \label{statesdef2}
& | 3^{\lambda_3} \,  4^{\lambda_4} \rangle \equiv    R(\phi, \frac{\theta}{2},-\phi)  | p_3\, S_3\, \lambda_3,\alpha_3 ; p_4\, S_4\, \lambda_4, \alpha_4 \rangle_{\substack{ \\ c.o.m. \\  \!\theta=\phi=0}} \,,
 \end{align} 
 while the S-matrix  is related to the (helicity) scattering amplitude of a $2\rightarrow 2$ process by 
\begin{equation}
 \label{defMmomentum}
\langle 3^{\lambda_3} \,  4^{\lambda_4} | S-\mathbb{I} | 1^{\lambda_1} \,  2^{\lambda_2} \rangle =(2\pi)^4\delta^4(p_{1}+p_2- p_{3}-p_4)\, i\, \mathcal{M}_{\lambda_1 \lambda_2}^{\lambda_3 \lambda_4}(p_i)\, .
 \end{equation}
The 2-particle states are decomposed on Poincar\'{e} irreps of definite angular momentum, leading to the partial waves decomposition of the scattering amplitude and the S-matrix given by
\begin{gather}
\label{partialwavessum} 
\mbox{Partial-waves}=\left\{ 
\begin{aligned}
 \mathcal{M}_{\lambda_1 \lambda_2}^{\lambda_3\lambda_4}(p_i) &=  \frac{\mathcal{N}}{2} e^{i(\lambda_{12}-\lambda_{34})\phi}  \sum_{\ell} (2\ell+1) d^{\ell}_{\lambda_{12} \lambda_{34}}(\theta)  {\mathcal{M}_\ell}^{\lambda_3 \lambda_4}_{\lambda_1\lambda_2}(s)  \\
{\mathcal{M}_\ell}^{\lambda_3 \lambda_4}_{\lambda_1\lambda_2}(s) &= \mathcal{N}^{-1}\int_{-1}^{1} \!\!d\!\cos\theta \,\, \,d^{\ell}_{\lambda_{12} \lambda_{34}}(\theta)  \mathcal{M}_{\lambda_1 \lambda_2}^{\lambda_3\lambda_4}(p_i)\big|_{\phi=0} \\
{S_\ell}^{\lambda_3 \lambda_4}_{\lambda_1\lambda_2}(s) &=  \mathcal{N}^{-1} \int_0^{2\pi}  \frac{d\phi}{2\pi}\int_{-1}^{1} \!\! d\!\cos\theta \,\,\,  d^{\ell}_{\lambda_{12} \lambda_{34}}(\theta)  \,  e^{-i(\lambda_{12}-\lambda_{34})\phi}   S_{\lambda_1 \lambda_2}^{\lambda_3\lambda_4}(p_i) 
\end{aligned}
\right.
\end{gather}
where the normalization is $\mathcal{N}=8\pi\sqrt{\frac{s}{|\boldsymbol{p}_1| |\boldsymbol{p}_3|}}$ and $\lambda_{ij}\equiv \lambda_i-\lambda_j$. The first two expressions are respectively the projection of the amplitudes on a complete basis of asymptotic states, with $d^\ell_{\lambda^\prime \lambda}(\theta) \equiv \langle \ell\, \lambda^\prime | e^{-i\theta J_2} | \ell \lambda \rangle$ the Wigner d-matrix, and its inverse expression for the expansion coefficients called partial waves $\mathcal{M}_\ell$. The $J_2$ is the rotation generator around the $y$-axis. The decomposition of the S-matrix is recovered by adding back the identity in \eqref{defMmomentum} suitably written in spherical coordinates, see \eqref{norm1change}. 

Notice that rotations of the scattering plane by an angle $\phi$ change the scattering matrix by an overall phase, \begin{equation}
\label{phiDependence}
\mathcal{M}_{\lambda_1 \lambda_2}^{\lambda_3\lambda_4}(p_i)=e^{i(\lambda_{12}-\lambda_{34})\phi} \mathcal{M}_{\lambda_1 \lambda_2}^{\lambda_3\lambda_4}(p_i)\big|_{\phi=0}\,,
\end{equation}
something we repeatedly use in the following by going back and forth from momentum to impact parameter space. 

In the following we often express the partial wave $S$-matrix \eqref{partialwavessum} via matrix exponentiation 
\begin{equation}
\label{SmatrixExp}
{S_\ell}^{\lambda_3 \lambda_4}_{\lambda_1\lambda_2}(s) =  \left(e^{2i\delta_\ell(s)}\right)^{\lambda_3 \lambda_4}_{\lambda_1\lambda_2}
\end{equation}
of the scattering phase matrix $\delta_\ell(s)$,  of matrix elements $\left(\delta_\ell\right)_i^j=\left(\delta_\ell\right)^{\lambda_3 \lambda_4}_{\lambda_1\lambda_2}$ where the collective indices  take values $i=(\lambda_1,\lambda_2)$  and $j=(\lambda_3,\lambda_4)$. The phase matrix, and in particular its large impact parameter expression, plays an important role in extracting physical observables, as we review in the following.

In order to simplify the notation we restrict hereafter to the case of $m_1=m_3$ and $m_2=m_4$.  
In this case, initial and final c.o.m. 3-momenta are equal, $|\boldsymbol{p}_1|=|\boldsymbol{p}_3|\equiv |\boldsymbol{p}|$, and 
\begin{equation}
\label{kinem}
\boldsymbol{p}_1\boldsymbol{p}_3=\boldsymbol{p}^2\cos\theta\,,\qquad \frac{1-\cos\theta}{2}=\sin^2\frac{\theta}{2}= \frac{\boldsymbol{q}^2}{4\boldsymbol{p}^2}\,,\qquad \boldsymbol{q}\equiv \boldsymbol{p}_3 - \boldsymbol{p}_1\,. 
\end{equation}
This kinematics is not necessarily elastic, as the states can still change spin $S$, helicity $\lambda$, internal quantum numbers etc.

\subsection{Eikonal as $SU(2)\rightarrow ISO(2)$ Contraction }
\label{contractionSection}
 Now that all notation is set up, we are ready to apply the eikonal limit \eqref{LimitContraction2} $\ell\to\infty$ with $\ell\theta\to$~fixed to the $S$-matrix expressed in partial waves \eqref{partialwavessum}.
We are interested in showing that the scaling limit \eqref{LimitContraction2} has a nice geometric implementation: it is captured by the group contraction $SU(2)\rightarrow ISO(2)$ where the isometries of a large sphere, $SU(2)\sim SO(3)$,  are well approximated by the isometries $ISO(2)$ of a tangent 2D euclidean plane.

 The goal is to study the limit \eqref{LimitContraction2} of the partial wave decomposition of the matrix elements \eqref{partialwavessum}, and we first focus on how this limit acts on the Wigner-$d$ matrix, by exploiting group theory arguments.
 Let us recall that the Wigner-$d$ matrix  in the partial wave decomposition are the matrix elements $d^\ell_{\lambda^\prime \lambda}(\theta)\equiv \langle \ell\, \lambda^\prime | e^{-i\theta J_2} | \ell \lambda \rangle$, where an unitary irrep of the rotation operator $R(0,\theta,0)=\mathrm{exp}(-i\theta J_2)$ acts on states of definite angular momentum, with a rotation around the $y$-axis. We remind that the generators of the algebra of rotations $SU(2)$ satisfy the following commutation relations and acts on states $| \ell \lambda\rangle$ as
\begin{align}
\label{su2}
SU(2):\qquad & [J_3,J_{\pm }]= \pm J_\pm\,, \qquad [J_{+},J_{-}]= 2 J_3\,,\qquad (J_{\pm}\equiv J_1\pm i J_2)\,, \\
\label{su2irrep}
 & J_3|\ell \lambda\rangle=\lambda |\ell \lambda\rangle\,, \qquad J_{\pm}| \ell \lambda \rangle= \sqrt{\mathcal{J}^2 -\lambda(\lambda\pm1)}| \ell \lambda\pm 1\rangle\,,
\end{align}
 where $\mathcal{J}^2=\ell(\ell+1)$ is the Caisimir operator, and the tower contains $2\ell+1$ states within the irrep, which are eigenstates of $J_3$ with $\lambda,\lambda^\prime=-\ell,\ldots, \ell$.
 
 When the total angular momentum becomes large, and in particular much larger than the helicity of the scattered particles $\lambda/\ell\ll 1$, the $SU(2)$ algebra \eqref{su2} contracts to 
 \begin{align}
ISO(2):\qquad & [j_3,j_{\pm }]= \pm j_\pm\,, \qquad [j_{+},j_{-}]= 0\,,\\
\label{iso2irrep}
 & j_3 | \lambda\rangle =\lambda |\lambda\rangle\,,\qquad j_{\pm} |\lambda \rangle = |\lambda\pm 1\rangle\, ,
\end{align}
where we defined $j_{\pm}\equiv  J_{\pm}/\sqrt{\ell(\ell+1)}$ and  $j_3\equiv J_3$. The resulting algebra, where the raising and lowering operators commute, and are both charged under $j_3$ rotations around the $z$-axis, is $ISO(2)$, i.e. the isometries of the 2D euclidean plane. As anticipated, this is unsurprising, because at large angular momentum we recover the flat-earth limit, where the isometries of a sphere reduce to those of a  plane.

As $ISO(2)$ is non-compact, there exist non-trivial infinite dimensional irreducible representations with $\lambda\in\mathbb{Z}$ or $\lambda\in\mathbb{Z}+1/2$. Given that the generator $J_2= \sqrt{\ell(\ell+1)}  (j_{+} - j_{-})/2i$ appears explicitly inside the Wigner $d$-matrix, it is convenient to work in a basis built by states which are simultaneously eigenstates of $j_+$ and $j_-$ (as they commute). The so-called continuous-spin basis $|\varphi\rangle$ perfectly serves this purpose as it satisfies 
\begin{equation}
\label{continSpinBasis1}
j_{\pm}|\varphi\rangle  = |\varphi\rangle e^{\mp i\varphi }\, ,
\end{equation}
where $\varphi$ is an angle. The continuous-spin basis and the $|\lambda\rangle$ basis are connected via a Fourier series, namely 
\begin{equation}
\label{continSpinBasis}
| \varphi \rangle \equiv \sum_{\lambda \in \mathrm{(half)integers}} e^{i\varphi\lambda} |\lambda \rangle\ \longleftrightarrow |\lambda\rangle =\int_{0}^{2\pi}\frac{d\varphi}{2\pi} e^{-i\lambda\varphi}|\varphi\rangle\,,
\end{equation}

In summary, in the large $\ell$ limit, the $|\ell \lambda \rangle$ states can be decomposed in a suitable basis of  $ISO(2)$ irreps for which $J_2$ matrix elements are diagonal.    This procedure allows us to recover the Wigner $d$-matrix $d^\ell(\theta)$ in the large angular momentum limit 
\begin{align}
\label{dlimitExplicit}
d^\ell_{\lambda^\prime \lambda}(\theta)   \xrightarrow[\ell\rightarrow \infty ]{\theta\rightarrow 0}    \int_{0}^{2\pi} \frac{d\varphi}{2\pi} e^{i(\lambda^\prime-
\lambda)\varphi} e^{i\theta \mathcal{J} \sin\varphi } =  J_{\lambda-\lambda^\prime}(\mathcal{J} \theta) \ , \qquad \mathcal{J}\equiv \sqrt{\ell(\ell+1)} 
\end{align}
which is just an integral representation of the Bessel $J_\nu(x)$ function\footnote{The function $J_2(x)$ should not be confused for the $SU(2)$ generator $J_2$. We recall  that $\lambda-\lambda^\prime$ is integer, and any $2\pi$-periodic integration range in \eqref{dlimitExplicit} is equivalent.} \cite{abramowitz+stegun}. Notice that this is a highly oscillating integral, whose matrix elements would average to zero, except for the region where $\theta \ell $ is finite. This is consistent with the regime of validity we were interested in,  \eqref{LimitContraction2}. In the next Sections we exploit this form of the Wigner-$d$ matrix, in order to recover the eikonal amplitude. 

Finally, here we focused on the case of 4 dimensions, however a similar contraction should hold for $SO(D-1)$ to $ISO(D-2)$ which we leave for future investigation.

\subsubsection{Impact Parameter Eikonal}\label{subsec:beikonal}
All the ingredients to recover \eqref{eikonalintro} have now been introduced. The large angular momentum limit of the Wigner-$d$ matrix discussed in \eqref{dlimitExplicit} is explicitly inserted in the scattering matrix decomposed in partial waves \eqref{partialwavessum}. Furthermore, we define the impact parameter 
\begin{equation}
\label{impactparamDef}
b\equiv \frac{\mathcal{J}}{|\boldsymbol{p}|} = \frac{\sqrt{\ell(\ell+1)}}{|\boldsymbol{p}|}\,,\qquad {\mathcal{M}_{\ell(b)}}^{\lambda_3 \lambda_4}_{\lambda_1\lambda_2}(s) \equiv {\mathcal{M}}^{\lambda_3 \lambda_4}_{\lambda_1\lambda_2}(s,b) \, ,
\end{equation}
where $b$ becomes a continuous parameter in the large $\ell$ limit.
We change variables in the partial waves decomposition to $\theta(|\boldsymbol{q}|)=|\boldsymbol{q}|/|\boldsymbol{p}| \left(1+O(\boldsymbol{q}/\boldsymbol{p})^2\right)$, and as the integral is dominated by the region $|\boldsymbol{q}|\ll |\boldsymbol{p}|$ we can safely extend the upper boundary of the integral to infinity. Lastly, we absorbed in the amplitude the helicity-dependent phase appearing in \eqref{dlimitExplicit}, by rotating it by an azimuthal angle according to \eqref{phiDependence}.
The partial wave decomposition then becomes
\begin{align}
\label{eikonalLimit2}
{\mathcal{M}}^{\lambda_3 \lambda_4}_{\lambda_1\lambda_2}(s,b) 
& \xrightarrow[b\rightarrow \infty ]{}   \frac{1}{ 8\pi}  \frac{1}{|\boldsymbol{p}| \sqrt{s} } \int_{0}^{\infty} \!\! d|\boldsymbol{q}||\boldsymbol{q}|  \int_{0}^{2\pi} \frac{d\varphi}{2\pi}  \,\, \,e^{i b |\boldsymbol{q}| \sin\varphi } \mathcal{M}_{\lambda_1 \lambda_2}^{\lambda_3\lambda_4}(|\boldsymbol{q}| \ll |\boldsymbol{p}|)\big|_{\substack{ \\ \\ \phi=\varphi}}\, .
\end{align}
Finally, let's recall that the amplitude in \eqref{eikonalLimit2} refers to $\boldsymbol{p}_1$ and $\boldsymbol{p}_3$ forming respectively an angle $\pm\theta/2$ w.r.t. the $z$-axis. The two vectors lie in a plane rotated by $\phi=\varphi$ w.r.t the $xz$-plane, implying that $\boldsymbol{q}=(q^1,q^2,0)=|\boldsymbol{q}|(\cos\varphi,\sin\varphi,0)$ is contained in the $xy$ plane. We define a 2D-vector $\mathbf{q}=|\boldsymbol{q}|(\cos\varphi,\sin\varphi)$ (in non-{\it italic} font) in the $xy$-plane.  We can thus interpret $b |\boldsymbol{q}| \sin\varphi$ as a scalar product $\mathbf{b}\mathbf{q}$ with  a 2D impact parameter vector $\mathbf{b}=b(0,1)$ forming an angle $\pi/2-\varphi$ with $\mathbf{q}$. This leads us to the following definition of {\it eikonal} or {\it impact parameter transform} 
\begin{equation}
\label{eikonalLimit3}
  {\mathcal{M}}^{\lambda_3 \lambda_4}_{\lambda_1\lambda_2}(\boldsymbol{p},\mathbf{b}) \equiv  \frac{1}{4|\boldsymbol{p}| \sqrt{s} } \int \frac{d^2\mathbf{q} }{(2\pi)^2} e^{i \mathbf{b} \mathbf{q} } \,\mathcal{M}_{\lambda_1 \lambda_2}^{\lambda_3\lambda_4} (\boldsymbol{p}, \boldsymbol{q})\big|_{\substack{ \boldsymbol{q}=(\mathbf{q},0)\\ \mathbf{q}^2\ll \boldsymbol{p}^2}}  \, .
\end{equation}
Up to normalization, this is just the Fourier transform that maps 2D-momentum $\mathbf{q}$-space to 2D impact parameter $\mathbf{b}$-space. 
  
 Because of \eqref{phiDependence} and the invariance of the 2D scalar product under 2D rotations, the large $\ell$-limit of the partial wave transform and the eikonal transform are unitarily equivalent \footnote{The  $\mathrm{exp}i(\lambda_{12}-\lambda_{34})\varphi$-factor is expected to show up for a $\varphi$-rotated scattering plane via \eqref{phiDependence}. The origin of the extra $-\pi/2$-factor in \eqref{phasesinMbis} can also be understood as following.  From the momentum-space eikonal expression \eqref{eikonalQspace} it follows that the net exchanged momentum $\boldsymbol{q}=2\partial_{\mathbf{b}}\delta(s,\mathbf{b})$ is aligned to $\mathbf{b}$ which thus lie in the scattering plane, while the total angular-momentum vector (in c.o.m. frame) $\boldsymbol{J}=\mathcal{J}\boldsymbol{p}_1\wedge \boldsymbol{p}_3/\boldsymbol{p}^2$ is transverse to the scattering plane and needs a $J_3$-rotation by $-\pi/2$ to make it aligned to $\mathbf{b}$. Equivalently, $\mathbf{b}$ is actually the 2D projection of the 3D vector $\boldsymbol{b}=\boldsymbol{J}\wedge \boldsymbol{p}/| \boldsymbol{p}|^2$ with $\boldsymbol{p}=(\boldsymbol{p}_1+\boldsymbol{p}_3)/2(\cos\theta/2)$, with $\cos\theta$ given by \eqref{kinem} and the normalization  fixed by \eqref{impactparamDef}. This explains why the $\ell\gg 1$ limit and eikonal amplitude are related by the little-group  rotation \eqref{phasesinMbis}.   }  
\begin{align}
\label{phasesinM}
&\mathcal{M}(\boldsymbol{p},\mathbf{b})=  U \mathcal{M}(s,b) U^\dagger \,, \\ 
\label{phasesinMbis}
& {\mathcal{M}}^{\lambda_3 \lambda_4}_{\lambda_1\lambda_2}(\boldsymbol{p},\mathbf{b})\big|_{\mathbf{b}=b(\cos\varphi,\sin\varphi)} = e^{i(\lambda_{12}-\lambda_{34})(\varphi -\pi/2)}{\mathcal{M}}^{\lambda_3 \lambda_4}_{\lambda_1\lambda_2}(s,b) \, ,
\end{align}
whenever $|\mathbf{b}|$ is set equal to $\mathcal{J}/|\boldsymbol{p}|$, given the definition \eqref{impactparamDef} ${\mathcal{M}_{\ell(b)}}(s)\equiv {\mathcal{M}}(s,b)$. The unitary transformation $U$~\footnote{Of matrix elements $U_{\lambda_i \lambda_j}^{\lambda_k \lambda_l}=\delta_{\lambda_i}^{\lambda_k} \delta_{\lambda_{j}}^{\lambda_{l}} e^{i\lambda_{ij}(\varphi-\pi/2)}$}  just inserts  little-group phases in the off-diagonal  terms,  without changing the eigenvalues which control the physical scattering angles and the time delays.
 
An equivalent expression in terms of the partial wave $S$-matrix $\eqref{SmatrixExp}$ and in an index-free notation via matrix exponentiation is 
\begin{equation}
\label{eikonalLimit4}
e^{2i\delta(s,\mathbf{b})}-\mathbb{I}=\frac{i}{4|\boldsymbol{p}| \sqrt{s} } \int \frac{d^2\mathbf{q} }{(2\pi)^2} e^{i \mathbf{b} \mathbf{q} } \mathcal{M}(\boldsymbol{p},\boldsymbol{q})\big|_{\substack{\boldsymbol{q}=(\mathbf{q},0) \\ \mathbf{q}^2\ll \boldsymbol{p}^2}}   
\end{equation}
where the matrix $\delta(s,\mathbf{b})$ is related to $\delta_{\ell(b)}(s)$ by the same unitary transformation \eqref{phasesinM} that connects the amplitude function of $b$ to the one function of $\mathbf{b}$, and $\mathbb{I}$ has matrix elements $\delta_{\lambda_1}^{ \lambda_3}\delta_{\lambda_2}^{\lambda_4}$.

So far no assumptions on the details on the theory were made and the only approximation used is the large partial wave limit. However, one must be consistent and check that the phase-shift is dominated by the contributions at large $b$. For instance, a theory of only contact terms would lead to delta functions localized at $b=0$, thus lying outside of the regime of validity of the Eikonal. In particular, the chosen theory must include non-analyticities in $q$, and the easiest way to achieve it is by including a massless mediator, such as a graviton or a photon. Moreover semi-classicality requires a large scattering phase that is achieved with transplanckian scattering or large charges in gravity and QED respectively. 
 
\subsubsection{Momentum-space Eikonal}
We discuss in this section the eikonal scattering amplitude in momentum space, that is the inverse relation of \eqref{eikonalLimit3} or \eqref{eikonalLimit4}.

Let's work with the partial wave amplitude \eqref{partialwavessum} specialized to $m_1=m_3$ and $m_2=m_4$ (hence $|\boldsymbol{p}_1|=|\boldsymbol{p}_3|=|\boldsymbol{p}|$) and $\phi=0$ 
\begin{equation}
\label{partialMomentumphi0}
\mathcal{M}_{\lambda_1 \lambda_2}^{\lambda_3\lambda_4}(p_i)\big|_{\substack{\\ \phi=0}}
= \frac{4\pi \sqrt{s}}{|\boldsymbol{p}| }   \sum_{\ell} (2\ell+1) d^{\ell}_{\lambda_{12} \lambda_{34}}(\theta)  {\mathcal{M}_\ell}^{\lambda_3 \lambda_4}_{\lambda_1\lambda_2}(s)\,.
\end{equation}
A first observation is that the Wigner-$d$ returns  $\delta_{\lambda_{12}  \lambda_{34}}$ as $\theta\rightarrow 0$, unless simultaneously $\ell\rightarrow \infty$. Therefore, the non-trivial $\theta$-dependence for $\theta\ll 1$ in  \eqref{partialMomentumphi0} comes again from the region of summation where $\ell\theta$ is finite. In that region we can use the limit  \eqref{dlimitExplicit} $d^\ell_{\lambda^\prime \lambda}(\theta)\rightarrow J_{\lambda-\lambda^\prime}(\mathcal{J}\theta)$ and approximate the series with an integral over the impact parameter  $b$ defined in \eqref{impactparamDef}, namely 
\begin{equation}
\label{eikonalQspaceLimit}
\mathcal{M}_{\lambda_1 \lambda_2}^{\lambda_3\lambda_4}(p_i)\big|_{\substack{ \\  \\ \phi=0}}  \longrightarrow   
4 |\boldsymbol{p}| \sqrt{s}  \int_0^\infty \!\!\! db b \int_{0}^{2\pi} d\varphi  e^{i(\lambda_{12}-\lambda_{34})\varphi} e^{i |\vec{q}|  |\vec{b}|\sin\varphi }   {\mathcal{M}}^{\lambda_3 \lambda_4}_{\lambda_1\lambda_2}(s,b) \, ,
\end{equation}
where we have also made the approximation $\theta(|\boldsymbol{q}|)=|\boldsymbol{q}|/|\boldsymbol{p}| \left(1+O(\boldsymbol{q}/\boldsymbol{p})^2\right)$ \footnote{Notice that $J_{\lambda-\lambda^\prime}(0)=\delta_{\lambda_{12}  \lambda_{34}}$ smoothly connectes to $d^\ell_{\lambda_{12}  \lambda_{34}}(0)=\delta_{\lambda_{12}  \lambda_{34}}$, even in the small-$\ell$ region, at small $\theta$.}. 
 Finally, using \eqref{phasesinMbis} in \eqref{eikonalQspaceLimit} and recalling that $\mathcal{M}(\boldsymbol{p}, \mathbf{b})$ is $2\pi$-periodic ($\lambda_{12}-\lambda_{34}$ is integer), we arrive at the momentum-space eikonal amplitude  
 \begin{gather}
\label{eikonalQspaceBis} 
\mbox{}\left\{ 
\begin{aligned}
\,\,&\mathcal{M}_{\lambda_1 \lambda_2}^{\lambda_3\lambda_4}(\boldsymbol{p},\boldsymbol{q})\big|_{\substack{\mathrm{eik}\\ \phi=0}}= 
 4 |\boldsymbol{p}| \sqrt{s}  \int \!\! d^2\mathbf{b}  \,\, e^{-i\mathbf{q} \mathbf{b} } \mathcal{M}_{\lambda_1 \lambda_2}^{\lambda_3 \lambda_4}
(s,\mathbf{b}) \, , \\
&\mathcal{M}_{\lambda_1 \lambda_2}^{\lambda_3\lambda_4}(\boldsymbol{p},\boldsymbol{q})\big|_{\substack{\mathrm{eik}\\ \phi=0}}= -i
 4 |\boldsymbol{p}| \sqrt{s}  \int \!\! d^2\mathbf{b}  \,\, e^{-i\mathbf{q} \mathbf{b} } \left(e^{2i\delta(s,\mathbf{b})}-\mathbb{I}\right)_{\lambda_1 \lambda_2}^{\lambda_3 \lambda_4} \, .
\end{aligned}
\right.
\end{gather}
We have used the subscript ``$\mathrm{eik}$'' to remind the regime of small $|\boldsymbol{q}|$ and large $\mathbf{b}$ that holds on each side respectively.

The \eqref{eikonalQspaceBis} refers to scattering in the $\phi=0$~plane where $\boldsymbol{q}=|\boldsymbol{q}|(1,0,0)$. The generic  $\phi\neq0$ case is given by the same expression, but with general $\boldsymbol{q}=|\boldsymbol{q}|(\cos\phi,\sin\phi,0)=(\mathbf{q},0)$, resulting in the usual helicity factor
\begin{equation}
\mathcal{M}_{\lambda_1 \lambda_2}^{\lambda_3\lambda_4}(\boldsymbol{p}, \boldsymbol{q})\big|_{\substack{\mathrm{eik}}}= e^{i(\lambda_{12}-\lambda_{34})\phi} \mathcal{M}_{\lambda_1 \lambda_2}^{\lambda_3\lambda_4}(\boldsymbol{p},\boldsymbol{q})\big|_{\substack{\mathrm{eik}\\ \phi=0}}\,, 
\end{equation}
 consistently with \eqref{phasesinMbis}.  

In summary, the momentum-space eikonal amplitude is in fact the 2D inverse-Fourier transform of the impact-parameter eikonal amplitude.

\subsection{The All-Order Eikonal Amplitude}
\label{Sec:allorders}
It is interesting to see what are the leading corrections to the eikonal amplitudes \eqref{eikonalLimit3} and \eqref{eikonalQspaceBis} that we have obtained using the large angular momentum limit of  $d^{\ell}(\theta)$, see \eqref{dlimitExplicit}. 
The Wigner-$d$ matrix can be expressed as well in terms of (Jacobi) polynomials in $\cos\theta$ (see  \eqref{DinJacobi}) which themselves admit a large-$\ell$ expansion in terms of Bessel functions. This limit is uniform in the whole interval for $\theta\in [0,\pi)$~\footnote{The interval of uniform convergence is open at $\theta=\pi$: a priori this would matter only for identical particles when one has a singularity  associated to backward scattering. However, even in that case, it's enough to just split the interval into two parts, and then use $d^\ell_{\lambda^\prime \lambda}(\pi-\theta)=(-1)^{\ell+\lambda^\prime}d^\ell_{\lambda^\prime -\lambda}(\theta)$ to extend the convergence  to $\theta=\pi$ included.} and has a known error from the truncation \cite{Szego,FrenzenWong,HoffmannScott}, leading to 
\begin{equation}
\label{dUniformlimit}
d^\ell_{\lambda^\prime \lambda}(\theta) =N_{\lambda^\prime,\lambda,\ell}\left(\frac{\theta}{\sin\theta}\right)^{1/2}J_{\lambda -\lambda^\prime}((\ell+\frac{1}{2})\theta)+\sqrt{\theta}O(1/\ell^{3/2})\,, 
\end{equation}
with a known (although not very informative) prefactor $N_{\lambda^\prime,\lambda,\ell}\rightarrow 1$ for $\lambda, \lambda^\prime/\ell\rightarrow 0$. 

The \eqref{dUniformlimit} shows that the limit of the Wigner $d$-matrix recovered in \sref{contractionSection} is indeed accurate, up to a relative order $O(\theta^2)$ in the large-$\ell$ limit. 
 This implies that the results we derived so far with the eikonal expansion  based on the limit $\ell\rightarrow\infty$ are valid in transplanckian scattering up to $O(\theta^2)$, which is 2PM order, included.~\footnote{Note that finite-$\ell$ corrections can be made $O(\sqrt{\theta}/\ell^{3/2})\sim \theta^2/(G s)^{3/2}\sim \lpl/b \times \lambda_s/b \ll 1$ just by the replacement $\sqrt{\ell(\ell+1)}\rightarrow \ell+1/2$.   }  \\
 
Since \eqref{dUniformlimit} does not require small angle, we can in fact extend the eikonal amplitude to all orders in $\theta$, in the large-$\ell$ limit.\footnote{An alternative approach, equivalent only in the semiclassical regime, would be using the large-$\ell$ expansion of the Bessel functions in \eqref{dUniformlimit} (see e.g. 9.2.1 of \cite{abramowitz+stegun}), which gives $
d^\ell_{\lambda^\prime \lambda}(\theta) \xrightarrow[\ell\theta \gg 1]{}   \frac{2}{\sqrt{(2\ell+1)\pi \sin\theta}}\cos((\ell+1/2)\theta-(\lambda-\lambda^\prime)\frac{\pi}{2}-\frac{\pi}{4})+O(1/\ell\theta)$, and then derive along the lines of \cite{Kol:2021jjc,Bautista:2021wfy}  an all-order radial action for the semiclassical scattering of all masses and spins. } 
This can be done directly by plugging the  uniform limit \eqref{dUniformlimit} into  the partial wave expansion \eqref{partialwavessum} that gives 
\begin{align}
\label{Eikonal3PMprim}
{\mathcal{M}}^{\lambda_3 \lambda_4}_{\lambda_1\lambda_2}(s,b)
=
 \frac{1}{ 8\pi}  \frac{1}{|\boldsymbol{p}| \sqrt{s} } \int_{0}^{\infty} \!\!\!\! d|\boldsymbol{q}||\boldsymbol{q}| \left(\frac{\theta(|\boldsymbol{q}|)}{\sin\theta(|\boldsymbol{q}|)}\right)^{1/2}  \!\! \!\!  J_{\lambda_{34}-\lambda_{12}}(b |\boldsymbol{p}| \theta(|\boldsymbol{q}|))  \mathcal{M}_{\lambda_1 \lambda_2}^{\lambda_3\lambda_4}(\boldsymbol{p},\boldsymbol{q})\big|_{\substack{ \\ \\ \!\! \phi=0}}
 \end{align}
up to an error $O(1/(bp)^{3/2})$,  where we have made the change of variables corresponding to \eqref{kinem}, and we have redefined the relation between $b$ and $\ell$ 
\begin{equation}
\mathcal{M}(s,b)\equiv \mathcal{M}_{\ell(b)}(s)\,,\qquad b\equiv (\ell+1/2)/ |\boldsymbol{p}| 
\end{equation} 
to match the partial wave amplitude. 
This can be interpreted as a 2D Fourier transform after the suitable change of variables $Q(|\boldsymbol{q}|)= |\boldsymbol{p}|\theta(|\boldsymbol{q}|)$ (or $|\boldsymbol{q}|(Q)=2|\boldsymbol{p}|\sin{Q/2|\boldsymbol{p}|}$), with a jacobian given by $|\boldsymbol{q}|d|\boldsymbol{q}|=|\boldsymbol{p}|\sin{\frac{Q}{|\boldsymbol{p}|}}dQ$,  where again $\theta(|\boldsymbol{q}|)$ is defined in \eqref{kinem}.

 After the change of variables, \eqref{Eikonal3PMprim} takes the form
\begin{align}
\label{Eikonal3PM}
{\mathcal{M}}^{\lambda_3 \lambda_4}_{\lambda_1\lambda_2}(s,b)
=
 \frac{1}{ 8\pi \sqrt{s}}  \int_{0}^{\infty} \!\!\!\! dQ\left(\frac{Q}{|\boldsymbol{p}|} \sin{\frac{Q}{|\boldsymbol{p}|}}\right)^{1/2}  \!\! \!\!  J_{\lambda_{34}-\lambda_{12}}(b \, Q)  \mathcal{M}_{\lambda_1 \lambda_2}^{\lambda_3\lambda_4}(\boldsymbol{p},\boldsymbol{q}(Q))\big|_{\substack{ \\ \\ \!\! \phi=0}}
 \end{align}
where $\boldsymbol{q}(Q)=|\boldsymbol{q}|(Q)(\cos\phi,\sin\phi,0)$ and  which, recalling the Bessel integral form and the relation \eqref{phasesinMbis} that relates $\boldsymbol{b}$ to $b$, leads us to the  {\it all-orders  eikonal amplitude} (in $\theta$)  
\begin{equation}
\label{allordersQ}
  {\mathcal{M}}^{\lambda_3 \lambda_4}_{\lambda_1\lambda_2}(\boldsymbol{p},\mathbf{b}) =  \frac{1}{4|\boldsymbol{p}| \sqrt{s} } \int \frac{d^2\mathbf{Q} }{(2\pi)^2} e^{i \mathbf{b} \mathbf{Q} } \, \widetilde{\mathcal{M}}_{\lambda_1 \lambda_2}^{\lambda_3\lambda_4} (\boldsymbol{p},Q) \qquad 
 \widetilde{\mathcal{M}}_{\lambda_1 \lambda_2}^{\lambda_3\lambda_4} (\boldsymbol{p}, Q) \equiv   \left(\frac{|\boldsymbol{p}|}{Q} \sin{\frac{Q}{|\boldsymbol{p}|}}\right)^{1/2}  \mathcal{M}_{\lambda_1 \lambda_2}^{\lambda_3\lambda_4}(\boldsymbol{p},\boldsymbol{q}(Q))
   \end{equation}
at a generic angle in the $xy$ plane. The \eqref{eikonalLimit3}  matches up to $O(\theta^2)$ the \eqref{allordersQ}.

 The eikonal amplitude in impact parameter space is recovered by applying $\int_0^\infty db b J_{\lambda_{34}-\lambda_{12}}(b\, Q^\prime)$ to both sides of \eref{Eikonal3PM} and using the orthogonality of the Bessel functions \eqref{BesselOrthogonality}. The resulting expression is
\begin{equation}
\mathcal{M}_{\lambda_1 \lambda_2}^{\lambda_3\lambda_4}(\boldsymbol{p},\boldsymbol{q})\big|_{\substack{ \\  \\ \!\! \phi=0}} =8\pi |\boldsymbol{p}|\sqrt{s} \left(\frac{\theta}{\sin\theta}\right)^{1/2} \int_0^\infty db b  J_{\lambda_{34}-\lambda_{12}}(b |\boldsymbol{p}| \theta)   {\mathcal{M}}^{\lambda_3 \lambda_4}_{\lambda_1\lambda_2}(s,b) \, ,
\end{equation}
where we reintroduced $\theta=Q^\prime/ |\boldsymbol{p}|$. The relative error decays in the semiclassical limit as $|\boldsymbol{p}|\rightarrow \infty$.

Using the Bessel-integral representation \eqref{dlimitExplicit}, and defining $\mathcal{M}(s, \mathbf{b})$ as in \eqref{phasesinMbis}, it leads to a generalized  {\it all-order  eikonal transform} in momentum space  \begin{gather}
\label{EikonalAllOrders} 
\mbox{}\left\{ 
\begin{aligned}
\,\,&\mathcal{M}_{\lambda_1 \lambda_2}^{\lambda_3\lambda_4}(\boldsymbol{p},\boldsymbol{q})\big|_{\substack{\boldsymbol{q}=(\mathbf{q},0) }} =4 |\boldsymbol{p}|\sqrt{s} \mathcal{N}(\theta) \int  \!\!  d^2 \mathbf{b}_e \,\, e^{-i \mathbf{b}_e \mathbf{q}}  {\mathcal{M}}^{\lambda_3 \lambda_4}_{\lambda_1\lambda_2}(s,\mathbf{b}(\mathbf{b}_e)) \, , \\
&\mathcal{M}_{\lambda_1 \lambda_2}^{\lambda_3\lambda_4}(\boldsymbol{p},\boldsymbol{q})\big|_{\substack{\boldsymbol{q}=(\mathbf{q},0)  }} =-i4 |\boldsymbol{p}|\sqrt{s} \mathcal{N}(\theta) \int  \!\!  d^2 \mathbf{b}_e \,\, e^{-i \mathbf{b}_e \mathbf{q}} \left(e^{2i\delta(s,\mathbf{b}(\mathbf{b}_e))}-\mathbb{I}\right)_{\lambda_1 \lambda_2}^{\lambda_3 \lambda_4}  \, .
\end{aligned}
\right.
\end{gather}
where  
\begin{align}
\label{expressionforN}
\mathcal{N}(\theta)=  \left[\left(\frac{\theta}{\sin\theta}\right)^{1/2}\left(\frac{\sin\theta/2}{\theta/2}\right)^2\right]=1+O(
\theta^4)\,, \qquad \mathbf{b}=  \left(\frac{\sin\theta/2}{\theta/2} \right) \mathbf{b}_e 
\end{align}
with the normalization $\mathcal{N}(\theta)$ becoming important only from the 4PM-order onward. 

The \eqref{EikonalAllOrders} is expressed in terms of the partial wave scattering phase $\delta_\ell$ by construction (hence enjoying all its properties, $\mathrm{Im}\delta_\ell \geq 0, \ldots$) and it delivers indeed the expected relation \eqref{TimeDelayScatteringAngle} between $\delta_\ell$ and the scattering angle\footnote{Explicitly, the stationary phase of \eqref{EikonalAllOrders} in the spinless case gives $
 \theta=\frac{2}{|\boldsymbol{p}|}\frac{\partial \mathrm{Re}\delta(s,\mathbf{b})}{\partial |\mathbf{b}|}= 2\frac{\partial \mathrm{Re}\delta_\ell(s)}{\partial \ell}$. For the expression valid away from the eikonal limit or for generic spins see \eqref{TimeDelayScatteringAngle2}: in summary, first diagonalize $\delta(s,\mathbf{b})$ then look for the saddle point.  } $\theta$ and time delay $\mathcal{T}$. It is however slightly different than the 2D Fourier transform of \cite{Ciafaloni:2014esa,DiVecchia:2021bdo} which, to our understanding, is a convenient definition of a suitable phase $\widetilde{\delta}$ made to match to the correct scattering angle.

\subsection{Scattering Angle and Time Delay}
\label{Sec:ScatAngleTimeDelay}

The scattering phase $\delta(s,\mathbf{b})$ is the master function from which we can extract other physical observables. Its partial derivatives, for instance,  are directly connected to the time-delay and scattering angle, as we show in this section. We first review how the observables emerge in the scattering spinless particles in \sref{Sec:timedelaynospin}, and then expand on \cite{Bellazzini:2021shn,Arkani-Hamed:2020blm,Maiani:1997pd} by adding the spin dependence as well in \sref{Sec:timedelayspin}. 

\subsubsection{Observables without Spin}
\label{Sec:timedelaynospin}

We consider a linear superposition of partial waves $ |\sqrt{s}\, \ell\, \alpha \rangle_{\mathrm{in}(\mathrm{out})}$ for an incoming (outgoing) 2-particle system in the c.o.m. frame of a given energy $\sqrt{s}$, total angular momentum $\ell$ and other internal quantum numbers $\alpha$
\begin{align}
 |f\rangle_{\mathrm{in/out}}= \int d\sqrt{s} \sum_{\ell, \alpha}f_{\ell,  \alpha}(\sqrt{s}) | \sqrt{s}, \ell, \alpha \rangle_{\mathrm{in/out}}\, . 
\end{align}
The wavepacket $f_{\ell,  \alpha}(\sqrt{s})$ is normalized to $\sum_{\ell \alpha}\int d\sqrt{s} |f_{\ell,  \alpha}(\sqrt{s})|^2=1$ and is sharply peaked around some particular center of mass energy $\sqrt{s}$, some $\ell$ and some $\alpha$. 

An outgoing 2-particle state differs at late time from an early-time incoming 2-particle state  by the action of the little-group associated to the vanishing 3-momentum of the c.o.m. frame ---i.e. 3D rotations--- or the action of time-translations which give rise to an overall phase, hence the same ray and physical observables.
Restricting to the non-trivial polar rotations, there should thus exist an angle $\theta$ (the scattering angle) and a time $\mathcal{T}$ (the time delay) for which the time-delayed and angle-rotated outgoing elastic state $e^{i\mathcal{T}H+i\theta J_2}|f \rangle_{\mathrm{out}}$ has the maximal overlap with the ingoing state $|f\rangle_{\mathrm{in}}$ in the semiclassical limit $\alpha_g\gg 1$ where the phase is large.   This overlap becomes
\begin{align}
& |{}_{\mathrm{out}}~\!\!\langle f| e^{-iT H-i\theta J_2} |f\rangle_{\mathrm{in}} |=| \sum_{\ell \alpha}\int d\sqrt{s} |f_{\ell, \alpha}(\sqrt{s})|^2 \mathrm{Exp}\left[i(2\delta_{\ell}(\sqrt{s})-\mathcal{T}\sqrt{s})\right]P_{\ell}(\cos\theta)|\, ,\nnl
 & \xrightarrow[l\rightarrow \infty ]{} | \sum_{ \alpha} \int d\sqrt{s}\, d\ell \, d\varphi |f_{\ell, \alpha}(\sqrt{s})|^2 \mathrm{Exp}\left[i(2\delta_{\ell}(\sqrt{s})-\mathcal{T}\sqrt{s}-\ell\theta \cos{\varphi})\right]|\, 
 \end{align}
where $P_{\ell}(\cos\theta)$ are the Legendre polynomial~\footnote{We have also replaced $\sum_{\ell} \leftrightarrow \int d\ell$,  but the statement is actually fully accurate thanks to the Poisson summation formula \cite{Berry:1972na,MorseFeshbach,Kol:2021jjc} which gives a deflection angle $\Theta(\theta)\equiv 2\pi n\pm \theta= 2\partial\mathrm{Re}\delta_{\ell, \alpha}/\partial( \ell+1/2)$ which differs by the net observed angle $\theta$ by an irrelevant  $2\pi n$,  where $n$ is the number of orbital winding between particles. Since the $\theta$ in this section is defined only up to $2\pi n$, we are allow to conflate $\theta$ with $\Theta$, which is effectively like just retaining the first winding mode. } 
. The phase-shift $\delta_{\ell}(\sqrt{s})$ has actually a real and imaginary part, the latter being entirely associated to particle production such as the emission of gravitons, photons or other states. The production of gravitons in the eikonal regime starts at 3PM.  
We have approximated the large $\ell$ behavior of the Legendre polynomials by the Bessel function $J_0(\ell \theta)$, which can be expressed in its integral form \eqref{dlimitExplicit}. 

The resulting integral in the limit \eqref{LimitContraction2} with $\alpha_g \gg 1$ (hence large $\delta_\ell$) is clearly controlled by two saddle  points: the time delay and the semi-classical deflection angle at
  \begin{equation}
  \label{TimeDelayScatteringAngle}
 \mathcal{T}=2\frac{\partial\mathrm{Re}\delta_{\ell}}{\partial \sqrt{s}}\,,\qquad \theta=2|\frac{\partial\mathrm{Re}\delta_{\ell}}{\partial\ell}|=\frac{2}{|\boldsymbol{p}|}\frac{\partial \mathrm{Re}\delta(s,\mathbf{b})}{\partial |\mathbf{b}|}\, .
 \end{equation}

\subsubsection{Observables with Spin}
\label{Sec:timedelayspin}
The same argument can be extended to the spinning case, where the partial waves $ |\sqrt{s}\, \ell\, \lambda, \alpha \rangle_{\mathrm{in}(\mathrm{out})}$ include total angular momentum $\ell$ of projection $\lambda$. The other internal quantum numbers are collectively called $\alpha=\{\lambda_i,\lambda_j,\ldots\}$, and include the helicities $\lambda_{i}$ of the single particles of which the state is made of. 
Let's specifically focus on linear superpositions 
\begin{equation}
| f \rangle_{\mathrm{in}(out)} = \int d\sqrt{s} \sum_{\ell \lambda \alpha \beta} f^\beta_{\ell,\lambda}(\sqrt{s}) V^{\alpha}_{\beta}  |\sqrt{s}\, \ell\, \lambda, \alpha \rangle_{\mathrm{in}(\mathrm{out})} \,,\qquad (V^\dagger S_\ell V)_{\alpha}^\beta=e^{2i\delta_{\ell,\alpha}}\delta_{\alpha}^\beta
\end{equation}
with wave-packets $f^\beta_{\ell, \lambda}(\sqrt{s})$ normalized to $\sum_{\ell \lambda \beta}\int d\sqrt{s} |f^\beta_{\ell, \lambda}(\sqrt{s})|^2=1$, and sharply peaked around some particular $\ell$, $\lambda$, $\alpha$ and $\sqrt{s}$, while $V$ is the unitary transformation acting on the internal indices that diagonalizes the partial-wave S-matrix ${S_\ell}_{\alpha}^{\beta}={}_{\mathrm{out}}~\!\!\langle \ell \lambda, \beta|  \ell \lambda, \alpha \rangle_{\mathrm{in}}$. 

 Following the spinless procedure, we study the overlap between the incoming and the time-delayed and angle-rotated outgoing elastic state
\begin{equation}
 |{}_{\mathrm{out}}~\!\!\langle f| e^{-iT H-i\theta J_2} |f\rangle_{\mathrm{in}} |=| \sum_{\ell \lambda \alpha}\int d\sqrt{s} |f^\alpha_{\ell, \lambda}(\sqrt{s})|^2 \mathrm{Exp}\left[i(2\delta_{\ell,\alpha}-\mathcal{T}\sqrt{s})\right]d^{\ell}_{\lambda\lambda}(\theta)|\, .
 \end{equation}
 For sharply-peaked wave-packets, and for large $\ell$ and $\sqrt{s}$, the overlap is controlled again by the stationary points of the semiclassical scattering phase which defines the (eigen-) scattering angles and time delays
  \begin{equation}
  \label{TimeDelayScatteringAngle2}
 \mathcal{T}_\alpha=2\frac{\partial\mathrm{Re}\delta_{\ell, \alpha}}{\partial \sqrt{s}}\,,\qquad \theta_\alpha=2|\frac{\partial\mathrm{Re}\delta_{\ell, \alpha}}{\partial \ell}|
 \end{equation}
where we have used \eqref{dUniformlimit} and the Bessel integral representation \eqref{dlimitExplicit}. The  \eqref{TimeDelayScatteringAngle2} gives a scattering angle and time delay for any eigen-scattering phase labeled by $\alpha$, generalizing the familiar results of \cite{Kol:2021jjc,Brandhuber:2021eyq,Damgaard:2021ipf} to the case of generic spin and inelasticity.  For example scattering of photons gives rise to two time-delays depending on the external helicity configuration. In summary, the observables in the spinning case are similiar to the scalar scenario, by simply replacing the scattering phase with its eigenvalues \footnote{While we focused on the most natural observables, the eigen-angles, one may be interested in less invariant information, such as the change of the scattering plane in certain spin configurations, etc. This can be extracted by a simple change of basis, from the  helcity-amplitude basis to a basis of spin pointing in certain   directions.  It would be perhaps interesting to make this change of basis explicit and make contact with e.g. \cite{Bautista:2021wfy}.}, $\delta \longrightarrow \delta^{\mathrm{diag}}$.

 The causality condition 
 \begin{equation}
 \label{causalitytimedel}
 \mathcal{T}\geq 0\, ,
 \end{equation}
 usually referred to as ``asymptotic causality'' is roughly the statement that  {\it interactions can only slow you down}. We actually prove \eqref{causalitytimedel} ---via analyticity and unitarity--- in  \sref{section:CausalityPositivity}, along with several other positivity/causality bounds.

\section{Transplanckian scattering}
\label{Sec:transplanckian}
In this section we consider  eikonal transplanckian scattering, $\alpha_g \gg 1$. 
The main object that we want to compute is the scattering phase. This is  a function of dimensionless quantities given by ratios of the scales of the problem introduced in \sref{sec:generalities}, namely 
\begin{equation}
\label{deltascales}
\delta=\delta \left(\alpha_g,\frac{ R_s}{b}, \frac{\lambdapl^2}{b^2}, \alpha,\frac{\lambdabar}{b}\right)\, .
\end{equation}

The leading term proportional just to $\alpha_g$ is called 1PM, while classical GR contributions scaling as $\left(R_s/b\right)^n$ and extracted from $(n-1)$-gravitational loops amplitudes are referred to as  $n$PM. Insertions of $G$ that give rise instead to $\lpl^2/b^2$-factors  are dubbed QG corrections, and we do not count them as genuine PM corrections, avoiding to put everything in the same basket.   We recall that $\alpha$ represents other couplings in the theory such as e.g. the electromagnetic fine structure constant, and $\lambdabar$ any other new length scale in the problem (e.g. the electron Compton wavelength or body finite size). 

A closer inspection reveals the following structure
\begin{equation}
\label{deltaPM}
\delta(s,\boldsymbol{b})=\alpha_g\sum_{m,n,p,q} \beta_{mnpq} \left(\frac{ R_s}{b}\right)^m\left( \frac{\lambda_{Pl}}{b}\right)^{2n} \alpha^p \left(\frac{\lambdabar}{b}\right)^q=\delta_0(s,\boldsymbol{b})+\delta_1(s,\boldsymbol{b})+...\, ,
\end{equation}
where $\alpha_g$ appears just as an overall prefactor, and $\delta_i(s,\boldsymbol{b})$ contains all contributions scaling as $G^{i+1}$, that is given by the sum over the $\beta_{mnpq}$ with $m+2n=i+1$.  The $m=n=p=q=0$ corresponds to the logarithmic  leading eikonal $\propto \alpha_g \log b$. Just to clarify the notation, the contributions to $\delta$ containing  no $(\lpl/b)$-factor are called 
\begin{equation}
(n+1)\mbox{PM}+p\mbox{Gauge} \longleftarrow (R_s/b)^n \alpha^p (\lambdabar/b)^q
\end{equation} 
without specifiying the $\lambdabar/b$-order that can change (as one varies $b$ across the scale $\lambdabar$), from  an analytic  contribution $(\lambdabar/b)^q$ (i.e. from local higher dimensional operators) to a non-analytic dependence such as $\log^2 \lambdabar/b$ due to light or mass particles  running in loops. 

The important observation at this point is that the dependence on the gravitational strength $\alpha_g$ is analytic, linear,  and that  the eikonal expansion holds for any value of $\alpha_g$ (provided  $\ell$ is large).  We are thus allowed to expand both sides of \eqref{eikonalLimit3} in the subplanckian eikonal regime, and extract the $\beta_{mnpq}$ from a perturbative calculation along the lines of  \cite{Amati:1990xe}.

Explicitly, the phase matrix at the first two orders in $G$ is thus given by
 \begin{gather}
\label{eikonalQspace} 
\mbox{}\left\{ 
\begin{aligned}
\,\,&\delta_0(s,\mathbf{b})= 
\frac{1}{8|\boldsymbol{p}| \sqrt{s} } \int \frac{d^2\mathbf{q} }{(2\pi)^2} e^{i \mathbf{b} \mathbf{q} } \,\mathcal{M}_{0} (\boldsymbol{p},\boldsymbol{q})\big|_{\substack{ \boldsymbol{q}=(\mathbf{q},0)\\ \mathbf{q}^2\ll \boldsymbol{p}^2}}  
 \, , \\
&\delta_1(s,\mathbf{b})+i\delta_0(s,\mathbf{b})^2= 
 \frac{1}{8|\boldsymbol{p}| \sqrt{s} } \int \frac{d^2\mathbf{q} }{(2\pi)^2} e^{i \mathbf{b} \mathbf{q} } \,\mathcal{M}_{1} (\boldsymbol{p},\boldsymbol{q})\big|_{\substack{ \boldsymbol{q}=(\mathbf{q},0)\\ \mathbf{q}^2\ll \boldsymbol{p}^2}}   \, .
\end{aligned}
\right.
\end{gather}
where we drop the helicity indices for notational simplicity.  The $\delta_0^2$ is intended as a product of matrices in helicity space. The amplitude $\mathcal{M}_i(\boldsymbol{p},\boldsymbol{q})$ includes all contributions scaling with $G^{i+1}$ or containing $i$ gravitational loops.
We recall that the leading eikonal phase at lowest order in all coupling is just $\delta^{\mathrm{lead}}_0(s,\mathbf{b})=-\alpha_g\log b/b_{\mathrm{IR}}$, where  $b_{\mathrm{IR}}$ is an IR cutoff specific to four dimensions, see Appendix~\ref{wordline-Eik} for a physical interpretation.

\subsection{Scattering Phase in QED+gravity}

Let's present now a fully worked-out example in the context of transplanckian scattering in  QED coupled to gravity, up to two-loop order.  More specifically, we consider the scattering of photon  off a neutral scalar at $s\gg \mpl^2$, $b\gg R_s$,  to  $O(G^2)$ and $O(\alpha)$, that is including loops of charged states as well as of gravitons.  We start first reviewing the results of \cite{Bellazzini:2021shn} at one loop, and then move to eikonal scattering to two loop order in \ref{sec:2PMapl}.

\subsubsection{1-loop: $O(\alpha)$ relative corrections}

In this subsection we are interested in extracting the relative $O(\alpha)$-corrections to the leading eikonal scattering phase $\delta^{\mathrm{lead}}_0(s,\mathbf{b})=-\alpha_g\log b/b_{\mathrm{IR}}$.  As we have seen in \sref{sec:classicalT}, these corrections may well dominate over the classical PM corrections as soon as $R_s$ is much smaller than the Compton wavelength of a charged state that couples significantly to the photon.  In particular, we focus on $\alpha_g \alpha\gg1$ with $1\ll (R_s/b)^2 \ll \alpha$ and $\lambdabar=1/m_e > R_s$, so that either massive or massless charged states in loops are more important than classical PM corrections. 
 Of course, this regime is never realized for astrophysical objects. It can be realized however in a huge range of values of $R_s/b$ when scattering SM particles or small black holes.\footnote{Example: tranplanckian scattering with $R_s=10^5\lpl$, while SM charged particles have Compton wavelengths a dozen orders of magnitude larger than $R_s$, and coupling to photon $O(\alpha_{\mathrm{em}})\sim 1/100$.  As expected, and confirmed in this section,  their quantum contribution to the phase shift can be tens of orders of magnitudes more important than classical 3PM corrections. } 
 
We consider the gravitational scattering of a photon against a neutral scalar particle that couples only gravitationally.  The exchanged graviton couples to the stress energy tensor of the photon, whose loop effects are encoded in form factors. The relevant amplitude in the $t/s\ll 1$ eikonal limit takes the form
\begin{align}\label{eik_amp}
\cM_0^{\mathrm{eik}}(s,t)= \alpha_g\frac{16\pi |\bm{p}|\sqrt{s}}{ \bm{q}^{\,2}}
\begin{pmatrix}
F_1(t) & -4 q_+^2F_3(t)  \\[0.3cm]
-4q_-^2 F_3(t)& F_1(t) 
\end{pmatrix}\ ,
\end{align}
where $q_{\pm}=\frac{1}{\sqrt{2}}(q_1\pm q_2)$, and the diagonal entries are helicity preserving while the off-diagonal describe helicity flipping. The details of the energy-momentum tensor form factors can be found in \cite{Bellazzini:2021shn}, and it depends on the species (e.g. the spin) of the particle running in the loop. The purely gravitational case ($\alpha=0$) is given by the diagonal matrix with $F_1(t)=1$, $F_3(t)=0$.

In \cite{Bellazzini:2021shn}, two main regimes are considered:  large mass of the charged states  where the particles can be integrated out leading to the effective $F^2R$ coupling in \eqref{eqExampleFFRie}, and the massless limit. The regimes correspond to probing the scattering at different impact parameters $b$ compared to the Compton wavelength of the state running in the loop. In particular, at $b\gg \lambdabar=1/m_e$, the corrections are due to a local operator in the EFT (associated to gravitational non-minimal coupling),  contributing with power-law $(\lambdabar/b)^2$  to the phase matrix 
\bea
\label{deltalargeblimit2}
\delta_\pm(s , \mathbf{b}\gg \lambdabar)=\alpha_g\left(-\log{b/b_{IR}} \pm \frac{8 F_3(0)}{b^2}\right)\, ,
\eea
where $F_3(0)=-c(\alpha/4\pi)1/m_e^2$ and $c$ as defined just below \eqref{eqExampleFFRie}. On the other hand, at $b\ll  \lambdabar$ the eigenstates of phase matrix are given by
\begin{equation}
\delta_{\pm}(s,\mathbf{b}\ll  \lambdabar)=\alpha_g\left(-\log{b/b_{IR}}-\frac{\beta_X}{g }\log{b m_e}^2\right)
\end{equation}
where $g$ is the gauge coupling, $m_e$ is the mass of state in the loop and $\beta_X$ is the beta function of charged spin-0 or spin-1/2 states. When running in loops are charged spin-1 states, there are some subtleties related to Sudakov resummations at exponentially small $b$, we refer to \cite{Bellazzini:2021shn} for a full IR-safe treatment of that case. As discussed in \eqref{Sec:timedelayspin}, the eigenvalues of the phase matrix are relevant to extract physical observables. 

The leading term $\log{b/b_{IR}}$ comes from the tree-level graviton exchange, it is spin independent and leads to the famous Shapiro time delay. 
Notice that all ``gravitational dependence'' is encoded in the overall coupling $\alpha_g$.
We observe no important difference on whether the scalar neutral particle in the scattering is massless or massive. This conclusion radically changes at 2PM, as we explore in the next Section.

\subsubsection{2-Loops: $O(\alpha \lpl^2/b^2)$ vs $O(\alpha R_s/b)$ relative corrections}
\label{sec:2PMapl}
As discussed in \sref{sec:classicalT}, as we move towards higher powers of $G$, we are only interested in contributions in $R_s/b$ as they are the only ones leading to resolvable effects, as opposed to QG $\lpl^2/b^2$ relative corrections that are never resolvable in the eikonal  transplanckian regime. In this Section we focus on the case of scattering a neutral {\it and massless} scalar off photons, and look for  corrections that can appear at $G^2$. 

We anticipate that in this massless case, in fact, only a non-resolvable contribution turns out to be produced (not only in the purely gravitational case but also at two loops including the $\alpha$-corrections). From this point of view, there is a sharp difference in nature between the massless and massive scenario at $G^2$, as in the latter larger resolvable effects are instead expected to appear. 

In the following, we proceed with the original approach of ACV, e.g. \cite{Amati:1990xe},  focusing on the s-channel cut  which allows to extract the imaginary part of $\delta_1(s,\mathbf{b})$ and then reconstruct the real part via a dispersion relation. We choose this perspective first to review and clarify the method (which was also used to compute the real part of $\delta_2(s,\mathbf{b})$ in \cite{DiVecchia:2021bdo}), and with the goal of extracting the gauge contribution at 2PM. 

The starting point is the expansion the exponential form of the S-matrix \eqref{eikonalQspace}, where we select the $G^2$-contributions and take the imaginary part in the s-channel
\begin{align}
\label{ImM1bdelta}
\operatorname{Im}_s \mathcal{M}_{1}(s,\mathbf{b})&=2s\left[\delta_0(s,\mathbf{b})^2+\operatorname{Im}_s \delta_1(s,\mathbf{b})\right]\nnl&=\operatorname{Im}_s \mathcal{M}_{1}(s,\mathbf{b})_{2-cut}+\operatorname{Im}_s \mathcal{M}_{1}(s,\mathbf{b})_{3-cut}\, .
\end{align}
Considering the 2PM with one insertion of $\alpha$ is equivalent generically to a two loop calculation, whose s-cuts can be organized in two- and three-particle cuts.


At large $b\gg \lambdabar$, the gauge contributions reduce to EFT corrections to the photon-graviton vertices:  therefore all $G^2$-contribution actually appears at one  loop, and the only diagram contributing to the imaginary part in the s-channel  is the box, obtained by gluing two on-shell amplitudes \eqref{eik_amp}. This is not the case at small $b$, where $G^2$ is obtained at 2 loops, and as such contributes to the $s$-channel discontinuity also with 3-particle cuts. 

\begin{figure}[t]
\centering
\includegraphics[width=0.3\linewidth]{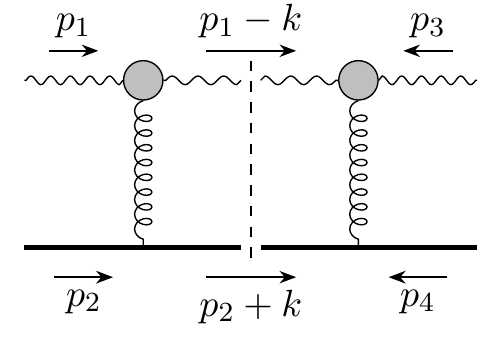}
\caption{{\small Pictorial representation of the R.H.S. of \eref{eq:ImM1}. In this picture, the grey blobs represent effective vertices of order $\alpha$. Recall that with in-coming momenta  $p_1^\mu + p_3^\mu = q^\mu$, and $s = (p_1 + p_2)^2$.}}
\label{fig:cutblob}
\end{figure}

We first analyse the two-particle cut, following and extending the argument of \cite{Ciafaloni:2014esa} to spinning states and loop corrections to the stress energy tensor. The cut is diagrammatically represented in \fref{fig:cutblob} and it is explicitely given by
\begin{align}\label{eq:ImM1}
\operatorname{Im}_s \mathcal{M}_{1}(s,t)_{2-cut}^{\lambda_1 \lambda_3}&=\frac{1}{2}\int {dk^4 \over (2\pi)^2}\mathcal{M}_0(s,k^2)_{\lambda_1 \lambda}\mathcal{M}_0((s,q-k)^2)\,_{\lambda \lambda_3} \, \delta_+((p_1-k)^2)\delta_+((p_2+k)^2)\nnl&={1\over 4s }\int {|k|d|k|d\phi \over (2\pi)^2}\mathcal{M}_0(s,k^2)_{\lambda_1 \lambda}\mathcal{M}_0(s,(q-k)^2)_{\lambda \lambda_3} \, ,
\end{align}
where the delta functions were solved explicitly in the center of mass frame of two incoming particles pointing in the z-direction as $\delta_+((p_1-k)^2)\delta_+((p_2+k)^2)={1\over 2 s |k|}\delta(k_0)\delta(\cos\theta-{|k|\over 2 E})$, and $\cos\theta$ is the projection of the loop momentum $k$ onto $p_1$. Since we are interested in the eikonal limit, the momentum exchanged by gravitons in the loop is small as selected by the saddle point at large $b$ (as further confirmed by \cite{Giddings:2011xs,Giddings:2010pp}), which means that we can directly use $\mathcal{M}_{0}\sim \mathcal{M}_{0}^{\mathrm{eik}}$ of Eq.~\eqref{eik_amp} in the limit of soft transferred momentum. As we want to compute $\delta_1$ at $O(\alpha)$, we can neglect the cross product $\alpha^2$. Furthermore, we shift the loop momentum in such a way that all gauge contribution is contained in $\mathcal{M}_0(s,k^2)$ (which amounts to multiply by 2 all terms proportional to $\alpha$) and $\mathcal{M}_0(s,(q-k)^2)=8\pi \alpha_g s/(q-Q)^2$ is simply the diagonal tree level gravitational term. The integral in $\phi$ is performed by a change of variable $z=e^{i\phi}$, where the integration over the unit circle is then given by the residue lying inside the contour, 
\begin{align}\label{eq:ImM2}
\operatorname{Im}_s \mathcal{M}_{1}(s,t)_{2-cut}^{\lambda_1 \lambda_3}&=\frac{-1}{4 s }\int \frac{|k|d|k| }{ 2\pi}\mathcal{M}_0(s,k^2)_{\lambda_1 \lambda_3}\frac{8\pi\alpha_gs}{ \sqrt{|q|^2+|k|^2}} +O(\alpha^2)\nnl
&={1\over 4s }\int {|\boldsymbol{k}|d\boldsymbol{k}d\phi' \over (2\pi)^2}\mathcal{M}_0(s,\boldsymbol{k}^2)_{\lambda_1 \lambda}\mathcal{M}_0(s,(\boldsymbol{q}-\boldsymbol{k})^2)_{\lambda \lambda_3} +O(\alpha^2)\, ,
\end{align}
where $|\boldsymbol{k}|=|k|$ and $|\boldsymbol{q}|=|q|$ are 2D vectors with angle $\phi'$ between them. The 4D cutted box at $O(\alpha)$ in the soft limit of the loop momentum becomes a 2D convolution \footnote{This property is observed also in the scattering of two heavy scalars, where the linearized delta functions can be trivially solved, see \cite{Herrmann:2021tct}} that factorized in $b$-space. We recover that the contribution of the two particle cut at $G^2$ and $\alpha$ is therefore 
\begin{equation}
\label{ImM1d0sq}
\operatorname{Im}_s \mathcal{M}_{1}(s,t)_{2-cut}^{\lambda_1 \lambda_3}=2s\left(\delta_0(s,\mathbf{b})^2\right)_{\lambda_1\lambda_3}\, ,
\end{equation}
where $\delta_0^2$ is intended as a product of matrices in helicity space. The immediate consequence of \eqref{ImM1d0sq} and \eqref{ImM1bdelta} is that $\operatorname{Im}_s\delta_1(s,\mathbf{b})=0$ at large $b\gg \lambdabar$, as there is no 3-particle cut contribution. On the other hand, let us anticipate that $\delta_1(s,\mathbf{b})$ develops an imaginary part at smaller impact parameters. This feature is completely equivalent to the 3PM term in pure gravity, where the effect of gravitational radiation appears due to the presence of a 3 particle cut of the H-diagram \cite{Amati:1990xe}, while here we encounter emission of SM particles already at $G^2$, as we shall analyse in a moment. 

Still within the $b\gg \lambdabar$ regime,  what can we say about the real part of $\delta_1(s,\mathbf{b})$? \\ One of the main features of the amplitude that can be exploited in order to reconstruct its real expression is crossing symmetry, which manifests itself simply as symmetry under $s\leftrightarrow u$ exchange as we are scattering a photon off a scalar spectator. As we have computed $\mathcal{M}_1$ already in the eikonal limit $t\ll s$, the properties of crossing symmetry are clearly lost, but it is still possible to reconstruct the real part by writing a basis of structures $s\leftrightarrow u$ symmetric, expanding them in the eikonal limit, and matching their imaginary part to the result previously computed in $b$-space. In practice, as $\delta_0$ is linear in $s$, we can select structures growing slower or equal to $s^3$. Furthermore, as we expect to produce a polynomial in s in the imaginary part, we can safely restrict to $\log$ type discontinuities.

We consider therefore the two following structures
\bea
s^3\log{(-s)}+u^3\log{(-u)}\sim s^3\left(-i\pi +{3 t\over s} \log{s}\right)+O(s)=X\, ,\nnl
s^2\log{(-s)}+u^2\log{(-u)}\sim s^2\left(-i \pi+2\log{s}\right)+O(s)=Y\, .
\eea
Notice that including higher powers of $\log{s}^n$ would lead to logarithms of $s$ in the imaginary part of $\mathcal{M}_1$, which is excluded by \eref{ImM1d0sq}. We expect $\mathcal{M}_1$ to be given by a linear combination of the two structures $\mathcal{M}_1=\alpha(t)X+\beta(t)Y$, but $\beta(t)=0$, as $\operatorname{Im}_s \mathcal{M}_{1}$ does not contain any $s^2$ contribution. By computing the Fourier transform and matching $\alpha(t)$ to \eqref{ImM1d0sq}, we can extract 
\bea
\delta_1(s,\mathbf{b})={3 \log{s} \over \pi s} \nabla^2_{\mathbf{b}} \delta_0(s,\mathbf{b})^2\, ,
\eea
which we remind is valid also at order $\alpha$ in the large $b$ regime. The scaling of this contribution is $\delta_1\sim \alpha_g (\lpl/b)^2$, which parallels the massless scalar result of ACV  \cite{Amati:1990xe}  here extended to photon-scalar scattering: \textit{no resolvable contribution is present at $O(G^2)$, that is no genuine $\alpha R_s/b$ relative correction  is found at this order for a massless neutral scalar scattering off a photon}. 

Obviously when more scales such as masses are involved in the problem the task of building a complete basis becomes much more involved. In \cite{DiVecchia:2021bdo}, the basis was extracted by fully computing the two loop contribution in $\mathcal{N}=8$, which turns out to be complete also for the GR scenario.
Another approach to recover $\delta_1$ is to compute the two particle cut without applying the small $t$ limit and then reconstruct the real part by using $s\leftrightarrow u$ dispersion relations. This task is easier when all external states are massive (and $t$ is small), as the $s$ and $u$ channel cuts are well separated, while extra care is needed when particles are massless and there is an overlap between the branch cuts. We leave this explicit approach to future work.

Is it surprising that only irrelevant non-resolvable QG corrections $\sim \alpha\lpl^2/b^2$  are found at $O(G^2)$ in the $b\gg \lambdabar$ limit? At this one loop order (see fig. \ref{fig:cutblob}) every amplitude can be decomposed in a basis of massless scalar amplitudes, which by direct inspection contains only logarithms. To be contrasted with the expected contribution at 2PM, which scales as $\frac{R_s}{b}=\frac{2G\sqrt{s}}{b}$, thus containing a square root type of discontinuity. The result is not surprising as it is consistent with this scaling argument.

\begin{figure}[t]
\centering
\includegraphics[width=0.8\linewidth]{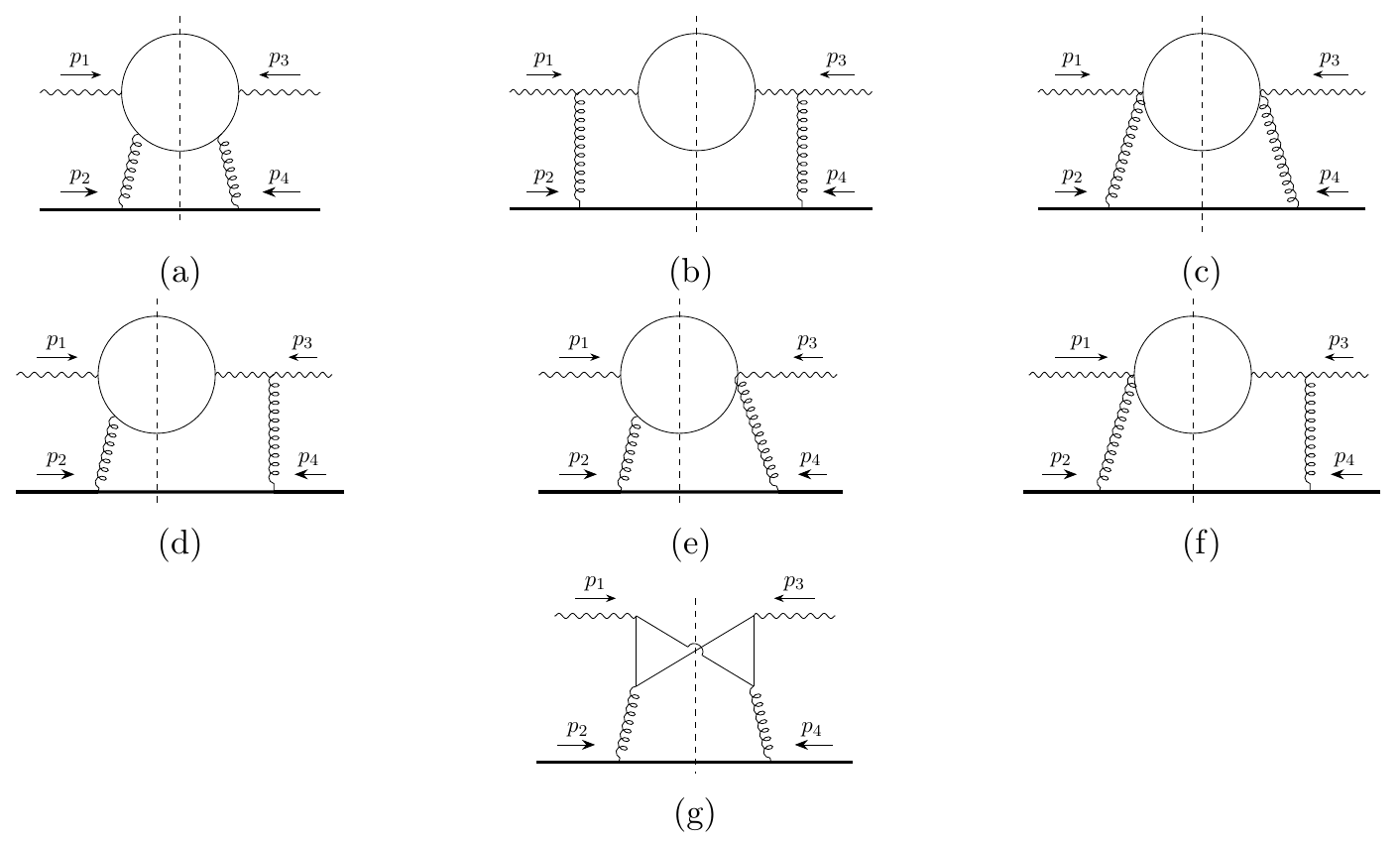}
\caption{{\small Topologies arising from the contraction ${\cal M}_{5\,\lambda_1}{\cal M}^*_{5\,\lambda_3}$. One must add the symmetric contributions w.r.t. the cut for (d), (e) and (f).}
\label{fig:topologies}}
\end{figure}

Let us now move to the study of the case  $b\ll \lambdabar$. As explained in the previous paragraphs, in this regime $\delta_1$ develops an imaginary part that we can extract by looking at the 3-particle cut of the amplitude. For simplicity, the charged particles in the loop are treated as effectively massless. The integrand of this computation is recovered by gluing two on-shell tree-level 5-point amplitudes ${\cal M}_{5\,\lambda_i}$ with one external photon. Considering all momenta incoming we have
\begin{align}
	\operatorname{Im}_s \mathcal{M}_{1}(s,t)_{3-cut}^{\lambda_1 \lambda_3}= \frac{1}{2}\int \dint{d}{k_1}\dint{d}{k_2}\dint{d}{k_3} \dd^{(d)}\big(p_1+p_2+k_1+k_2+k_3\big) \dd_+(k_1^2)\dd_+(k_2^2)\dd_+(k_3^2) {\cal M}_{5\,\lambda_1}{\cal M}^*_{5\,\lambda_3} \, ,
	\label{eq:2loop-3c}
\end{align}
where, to keep track of possible divergence terms, we work in $d = 4 -2 \varepsilon$ dimensions. We have also introduced the shorthand notation $\dd^{(n)}(x) \equiv (2\pi)^n \delta(x)$ for convenience.
The product of the two five-point amplitudes can be organise in terms of the six planar and one non-planar topologies represented respectively in \fref{fig:topologies}(a) -- (f) and \fref{fig:topologies}(g). To solve the two-loop integral given in eq.~\eqref{eq:2loop-3c}, we follow a procedure similar to the one employed in \cite{Parra-Martinez:2020dzs,DiVecchia:2021bdo,Herrmann:2021tct}. First of all, we parametrize the external kinematic as
\begin{align}
	p_1^\mu = \bar{p}^\mu_1 + \frac{q^\mu}{2} \, , & & p_3^\mu = - \bigg(\bar{p}^\mu_1 - \frac{q^\mu}{2} \bigg) \, , & &  p_2^\mu = \bar{p}^\mu_2 - \frac{q^\mu}{2} \, , & & p_4^\mu = - \bigg(\bar{p}^\mu_2 + \frac{q^\mu}{2} \bigg) \, .
	\label{eq:pbardef}
\end{align}
As a consequence, one can see that
\begin{align}
	\bar{p}_1 \cdot q = 0 = \bar{p}_2 \cdot q \, , & & \bar{p}_1^2 = -\frac{q^2}{4} = \bar{p}_2^2 \, .
\end{align}
Notice that w.r.t. Refs. \cite{DiVecchia:2021bdo,Herrmann:2021tct,Parra-Martinez:2020dzs} we do not expand the integrand in small $q^2$.

\begin{figure}[t]
\centering
\includegraphics[width=0.8\linewidth]{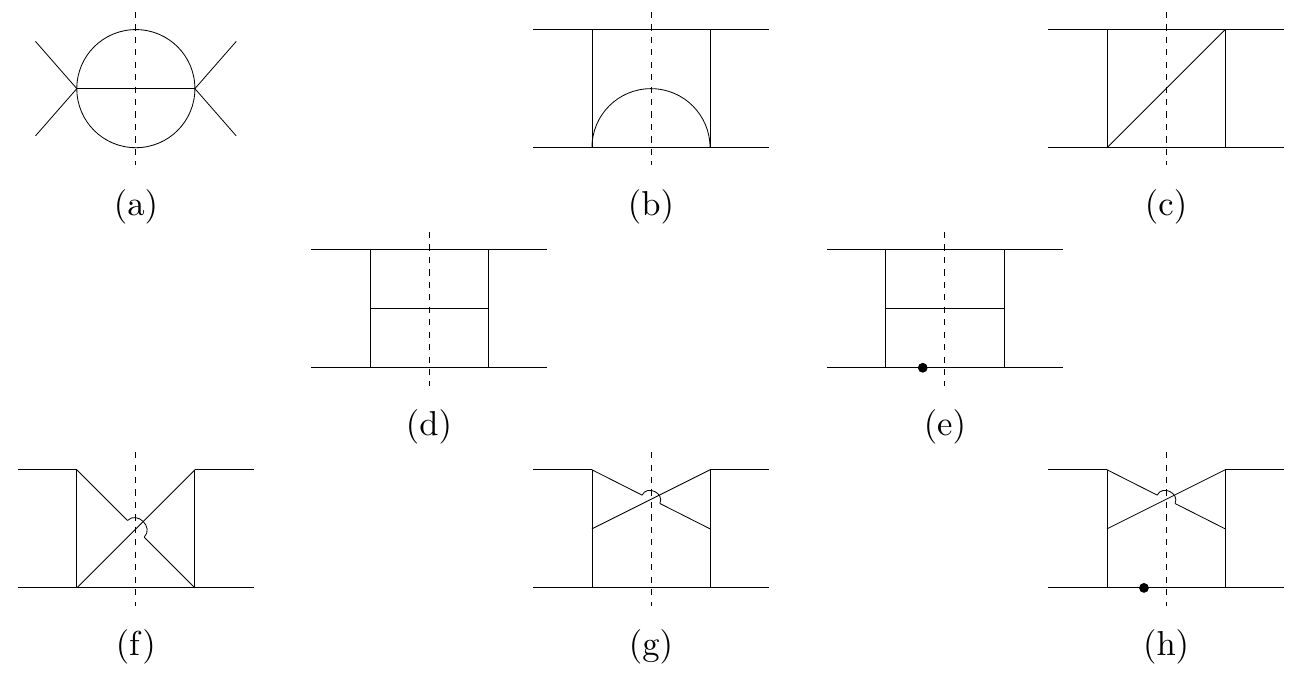}
\caption{\small{ The eight MIs needed to solve the two-loop integral. (a) -- (e) are the planar MIs, while (a) -- (c) and (f) --(h) are the non planar one. A dot on the propagator line means that it is squared in the integral.}}
\label{fig:MIs}
\end{figure}

Then, making use of reverse unitarity \cite{Anastasiou:2002yz,Anastasiou:2002qz,Anastasiou:2003yy,Anastasiou:2015yha}, we can  apply Integration-by-Parts (IBP) identities \cite{Tkachov:1981wb,Chetyrkin:1981qh, Smirnov:2012gma} with the help of the LiteRed package \cite{Lee:2012cn,Lee:2013mka} o rewrite eq.~\eqref{eq:2loop-3c} as a combination of two-loop scalar Master Integrals (MIs) sketched in \fref{fig:MIs}. In particular, the sum of the planar topologies can be written entirely in terms of five MIs depicted in \fref{fig:MIs}(a) --(e) that we can call $\vec{g} =\{g_1, g_2, g_3, g_4, g_5\}$. The non planar topology can be reduced to a combination of six master integrals depicted in  \fref{fig:MIs}(a) -- (c) and (f) -- (h), from now on called $\vec{\tilde{g}} =\{\tilde{g}_1, \tilde{g}_2, \tilde{g}_3, \tilde{g}_4, \tilde{g}_5, \tilde{g}_6\}$. Their explicit expression is written in \aref{app:MIs}. Schematically we have
\begin{equation}
	\operatorname{Im}_s \mathcal{M}_{1}(s,t)_{3-cut}^{\lambda_1 \lambda_3} = G^2 \alpha(-q^2)^2 \left(\sum_{j=1}^5 {\cal C}_j(y, \varepsilon) g_j(y,\varepsilon) + \sum_{j=1}^6 \tilde{{\cal C}}_j(y, \varepsilon) \tilde{g}_j(y,\varepsilon)\right) \, ,
	\label{eq:schematicMsol}
\end{equation}
where ${\cal C}_j$ and $\tilde{{\cal C}}_j$ are some polynomial functions of $\varepsilon$ and $y$. Notice that $\tilde{g}_1,\tilde{g}_2$ and $\tilde{g}_3$ are essentially equal to $g_1, g_2$ and $g_3$, and that $\tilde{g}_4$ is related to $g_3$ via crossing. Hence, in the end we have seven MIs to compute.

One can realise that the integrals $\vec{g}$ and $\vec{\tilde{g}}$ are dimensionless scalar functions and depends non-trivially only on the dimensionally regularisation parameter $\varepsilon$ and the variable
\begin{equation}
	y \equiv \frac{\bar{p}_1\cdot\bar{p}_2}{|\bar{p}_1||\bar{p}_2|} = \frac{2 s}{-q^2}-1 \, .
\end{equation}
Using the approach of Refs.~\cite{Kotikov:1990kg,Kotikov:1991pm,Bern:1992em,Gehrmann:1999as,Henn:2013pwa,Caron-Huot:2014lda}, we can solve these MIs by writing a suitable differential equation. These equations have been written in a canonical form using the package Fuchsia \cite{Gituliar:2016vfa,Gituliar:2017vzm}. In particular, the planar and non planar MIs satisfy respectively the following differential equations
\begin{align}
	d \vec{g}(\varepsilon, y) & = \varepsilon \left[ A_{+1} d \log(y+1) + A_{-1} d \log(y-1)\right]\vec{g}(\varepsilon, y)\, , \label{eq:diffplanar}\\
	d \vec{\tilde{g}}(\varepsilon, y) & = \varepsilon \left[ \tilde{A}_{+1} d \log(y+1) + \tilde{A}_{-1} d \log(y-1)\right] \vec{\tilde{g}}(\varepsilon, y) \, ,
	 \label{eq:diffnonplanar}
\end{align}
where $A_{\pm 1}$ and $\tilde{A}_{\pm 1}$ are constant matrices whose explicit expressions are written in \aref{app:MIs}. Expanding the integrals in $\varepsilon$
\begin{equation}
	g_i(\varepsilon, y) = \sum_n \varepsilon^n g_i^{(n)}(y) \, , \qquad \tilde{g}_i(\varepsilon, y) = \sum_n \varepsilon^n \tilde{g}_i^{(n)}(y) \, ,
\end{equation} 
one can easily solve the previous differential equations order per order in $\varepsilon$. 

In order to fix a unique solution, we just need to know the value of the MIs for a certain value of $y$, e.g. $y=1$. To this end, we can first solve the differential eqs. \eqref{eq:diffplanar} and \eqref{eq:diffnonplanar} around $y=1$ by essentially exponentiating the matrices $A_{-1}$ and $\tilde{A}_{-1}$. Then, following Refs. \cite{Caron-Huot:2014lda,Henn:2014qga}, we can use the fact that our basis of MIs is UV finite, hence study the behaviour of this solution for $\varepsilon < 0$ and require  its regularity. With this manipulation, we find some non-trivial relations between the various MIs in $y=1$. In particular, we find that $g_3(y=1)=0$ and that it is enough to compute the values of $g_1, g_4$ and $\tilde{g}_5$ in $y=1$ to uniquely fix all the boundary conditions for eqs. \eqref{eq:diffplanar} and \eqref{eq:diffnonplanar}. These integrals are either easy to compute or known in the literature, see e.g. \cite{Smirnov:1999gc,Tausk:1999vh}.

Regardless of this, one can realise that all the MIs are just (poly)logarithmic function of $y$. Therefore, the scaling of $\operatorname{Im}_s \mathcal{M}_{1}(s,\mathbf{b})_{\lambda_1 \lambda_3}$ is fixed by the polynomial in front of  $\vec{g}$ and $\vec{\tilde{g}}$ in \eref{eq:schematicMsol}, i.e. 
\begin{equation}
	\operatorname{Im}_s \mathcal{M}_{1}(s,t)_{3-cut}^{\lambda_1 \lambda_3} \sim G^2 \alpha q^4 y^n \sim \alpha_g \left( G \alpha \frac{s^{n-1}}{(q^2)^{n-2}} \right) \, ,
\end{equation}
where $n$ is an integer number smaller than 4. From this simple scaling reasoning, we can then see that no PM classical contribution proportional to $2G \sqrt{s} q \sim R_s/b$ is actually generated. Therefore, we find no resolvable effect at this order,  for the case of scattering of a photon off a massless neutral scalar.

\section{Analyticity, Causality and Time Delay}
\label{section:CausalityPositivity}

In this section we study the causal structure of eikonal amplitudes. At the operational level we are working with scattering amplitudes so that we trade causality for  suitable regions of analyticity in the external momenta or Mandelstam invariants. This can either be an assumption or it can be justified by  micro-causality (which holds also in fixed curved spacetimes  \cite{Dubovsky:2007ac}) together with angular momentum selection rules that fix the little-group scaling and the kinematical singularities associated.   

 We assume these analyticity properties apply as well to eikonal gravitational scattering because we are systematically neglecting all quantum gravity $O(\lpl/b)^n$-effects in this regime, while retaining instead  PM and QFT effects. 
This is closely related to  QFT on a fixed background except that we include radiation effects from gravity being dynamical, i.e. with the graviton being a state that can be produced in the scattering and put on-shell, and as such it enters in the imaginary part of the elastic amplitudes. 

As case of study, we focus on the instructive graviton-scalar scattering $1_{h}  2_S \rightarrow 3_{h} 4_S$   which {\it i)} retains a rather simple but non-trivial kinematic structure allowing  to isolate the dynamical singularities from the kinematical ones,  while  {\it ii)} it provides rather neat and interesting causality bounds with potential implication on modified-gravity theories, see also \cite{Serra:2022pzl}. 

Our analysis below shows the emergence of  a rich structure of causality bounds in the eikonal regime, in the form of non-linear positivity constraints, which largely extends the early on results of \cite{Camanho:2014apa,Adams:2006sv}, with implications and techniques similar to those studied recently in  e.g. \cite{Bellazzini:2020cot,Caron-Huot:2022ugt,Arkani-Hamed:2020blm,Bern:2021ppb} by introducing dispersive ``arcs'' in the complex $s$-plane, see Fig.~\ref{fig:arcs}.  We essentially parallel \cite{Caron-Huot:2022ugt} in proving positivity of the time-delay matrix, here for  particles with different spins and masses, and with moreover potential phenomenological implications in cosmology.  Furthermore, we find that eikonal scattering can be used to bound  even contact operators by inserting them into loops that correct the high-energy scaling of the scattering phase matrix.

\subsection{Graviton-Scalar dispersion relation}
 
 The gravitational graviton-scalar amplitude  can be written exposing the little-group dependence which, in the all-incoming momenta convention, takes in the most general case the following form 
 \begin{align}
\label{GravitonCompton}
\mathcal{M}(1_{\lambda_{1}}, 2\,, 3_{\lambda_3}, 4)=
\begin{pmatrix}
 \la 3  k_2 1]^4 F_{+-}(s,t)  &[ 13 ]^4  F_{++}(s,t)  \\
\la 13 \ra^4  F_{--}(s,t) &   \la 1 k_2 3]^4 F_{+-}(u,t)   
\end{pmatrix}_{\lambda_1 \lambda_3} \equiv \mathcal{M}_{\lambda_1 \lambda_3}(s,t) \equiv \mathcal{M}_{\lambda_1}^{-\lambda_3}(s,t)\, ,
\end{align}
with lower index helicities $(\lambda_1,\lambda_3)=(\pm2, \mp2)$ being helicity-preserving scattering, and $(\lambda_1,\lambda_3)=(\pm 2, \pm2)$  helicity-flipping. 
 We have  introduced the formal notation $\mathcal{M}_{\lambda_1}^{-\lambda_3}(s,t)=\mathcal{M}(1^{\lambda_1} 2 \longrightarrow 3^{-\lambda_3} 4)=\mathcal{M}(1_{\lambda_{1}}, 2\,, 3_{\lambda_3}, 4)=\mathcal{M}_{\lambda_1 \lambda_3}(s,t)$.

By Bose symmetry, crossing, and neutrality of the scalar and graviton, it follows that $F_{+-}(s,t)=F_{-+}(u,t)$, $F_{++}(s,t)=F_{++}(u,t)$, and $F_{--}(s,t)=F_{--}(u,t)$ \footnote{This would restricts further  to two independent form factors if one were to demand time-reversal or parity.}.
The tree-level values are proportional to 
\begin{equation}
F_{+-}(s,t)\big|_{\mathrm{tree}} \propto  \frac{G}{t(u-m^2)(s-m^2)}\,, \qquad F_{++}(s,t)\big|_{\mathrm{tree}} \propto \frac{Gm^4}{t(u-m^2)(s-m^2)}\, .
\end{equation}
 The form factors $F_{ij}(s,t)$ are free of any kinematical singularity in both $s$ and $t$, in agreement with \cite{Gross:1968in}, and ready-to-use for {\it unsubtracted}  dispersive relations thanks to the superconvergence  \cite{Abarbanel:1967wk,Gross:1968in},  as provided by the little group factors.  
 
 We could work directly with the form factors in the following, but for the purpose of studying causality constraints in the eikonal limit it is actually more convenient to deal directly with the eikonal amplitude. 
In fact, since  the little-group factors are
 \begin{equation}
  \la 3  k_2 1]^4= \la 1 k_2 3]^4=(su-m^4)^2\,, \qquad [ 13 ]^4=t^2 e^{4i\phi}\,, \qquad \la 13 \ra^4=t^2 e^{-4i\phi}\, ,
  \end{equation}
 the analytic properties of form factors and amplitude are actually the same (at $\phi=0$) in the complex $s$-plane at fixed (and negative) $t$. Hence, we can run the familiar dispersive arguments directly on the amplitude matrix elements.  

   \begin{figure}[t]
\centering
\includegraphics[width=0.4\linewidth]{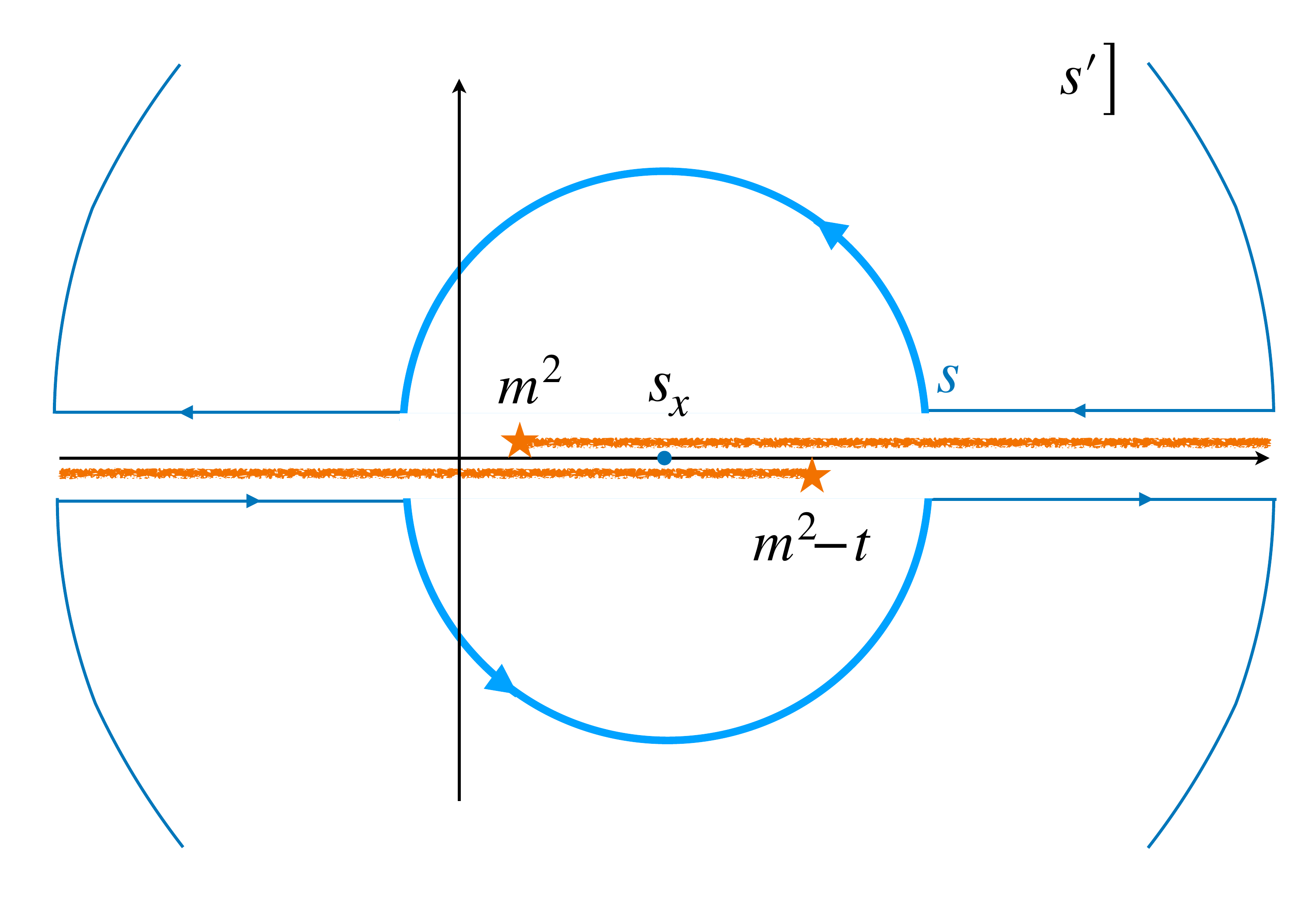}
\caption{{\small Thick blue lines represent the contour integral defining the arc \eqref{eq:arcDef}. Lighter blue lines correspond to the contour deformation giving rise to the UV representation \eqref{arcDisc1}. Orange lines on the real axis represent $s$ and $u$ channel branch-cuts. }}
\label{fig:arcs}
\end{figure}

We define thus the following  ``arcs''  as the contour integral over two half circles in the upper and lower $s^\prime$-plane centered at the $s-u$ crossing symmetric point $s_\times(t)\equiv m^2-t/2$
\begin{equation}
\label{eq:arcDef}
a^{(n)}_{\lambda_1 \lambda_3 }(s,t) \equiv \oint \frac{ds^\prime}{2\pi i } \frac{\mathcal{M}_{\lambda_1 \lambda_3}(s^\prime,t) }{[(s^\prime-u^\prime)/2]^{2n+3}} \, ,
\end{equation}
and with  radius $s-s_\times(t)>t/2$ \footnote{In the physical region, the $s$ and $u$-branch cuts are superposed when $ m^2<|s|<m^2-t$, so we choose the arc large enough to avoid that region.}. We have $u^\prime=u(s^\prime)=2m^2-s^\prime-t$. We can deform their contour along the real axis as shown in Fig.~\ref{fig:arcs}
\begin{equation}
\label{arcDisc1}
a^{(n)}_{\lambda_1 \lambda_3}(s,t) = \frac{2}{2\pi i} \int_{s}^{\infty} \!\!\!ds^\prime\frac{\mathrm{Disc}\mathcal{M}_{\lambda_1 \lambda_3}(s^\prime,t)}{(s^\prime-s_\times(t))^{2n+3}}\,, 
 \end{equation}
where  we recall that $s>s_\times+\frac{t}{2} $. We have used crossing symmetry to integrate only over a single physical region, $12\rightarrow 34$, and defined $\mathrm{Disc}\mathcal{M}_{\lambda_1 \lambda_3}(s,t)\equiv \mathcal{M}_{\lambda_1 \lambda_3}(s+i\epsilon,t)-\mathcal{M}_{\lambda_1 \lambda_3}(s-i\epsilon,t)$. 
By hermitian analyticity the discontinuity is nothing but the matrix element 
\begin{equation}
\langle  3^{\lambda_3} \,  4  | \mathcal{M}-\mathcal{M}^\dagger | 1^{\lambda_1} \,  2 \rangle= i \langle  3^{\lambda_3} \,  4  | \mathcal{M}^\dagger \mathcal{M} | 1^{\lambda_1} \,  2 \rangle
\end{equation}
so that the optical theorem in the forward elastic limit implies that both $\mathrm{Disc}\mathcal{M}_{+-}(s,t\rightarrow0)$ and $\mathrm{Disc}\mathcal{M}_{-+}(s,t\rightarrow 0)$ are positive\footnote{We adopt the following notation for simplicity: $\mathcal{M}_{-+}\equiv \mathcal{M}_{-2,+2}$,  $a^{(n)}_{+-}\equiv a^{(n)}_{+2,-2}$ etc.}, hence  the positivity bounds 
\begin{equation} 
\label{Arcelasticbounds}
a^{(n)}_{+-}(s,t\rightarrow 0)>0\,, \qquad a^{(n)}_{-+}(s, t \rightarrow 0)>0\,. 
\end{equation}
These two bounds hold as long as the dispersive integrals are convergent  ($n\geq 0$, due to the Regge bound \cite{Haring:2022cyf}), and the forward limit exists ($n\geq 1$, due to the $t$-channel pole). 

 For example, the $n=1$ bounds imply $F_{+-}(0,0)>0$  whenever IR loops are negligible. This corresponds to a tree-level bound on contact operators in scalar-tensor theories. For instance, at the 8th order in derivatives there are 3 independent operators in a shift-symmetric and parity-preserving scalar-tensor theory \cite{Ruhdorfer:2019qmk}, but only one enters in $F_{+-}(0,0)$,  that can be choosen e.g. to be  $\delta \mathcal{L}_8= \xi/\Lambda^6 (\nabla_\mu \nabla_\nu \phi R^{\mu\alpha\nu\beta})^2$.  Therefore,  the $n=1$ tree-level bound is  $\xi>0$.  In this example, the IR loops are negligible as long as $\Lambda\ll \mpl$ and the insertions of other operators such as e.g. $\tilde{\xi}(\partial\phi)^2 (\nabla\nabla\phi)^2/\Lambda^6$ into loops are not enhanced, i.e. $\tilde{\xi}/\xi \ll 16\pi^2$.    Of course, IR loops in an  EFT are calculable and they can always be included systematically if needed, the sharp statement being \eqref{Arcelasticbounds}. 
 Notice, moreover, that the $s$-derivative of the forward arcs in \eqref{arcDisc1} is negative definite, $\partial_s a^{(n)}(s,t=0)<0$, so that IR loops make the arcs larger as $s$ is taken smaller. 
  
More general positivity conditions can be extracted by exploiting the partial wave expansion  \eqref{partialwavessum} with $\phi=0$ that makes the unitarity condition block-diagonal in $\ell$, that is~\footnote{We have resorted to a lighter notation where inessential labels, such as the 4-momentum are suppressed. We have also removed the trivial 4-momentum conservation delta-function.} 
\begin{align}
\label{imaginarypartMatrix}
\frac{1}{2i}\mathrm{Disc} {\mathcal{M}}_{\lambda_1 \lambda_3}(s>s_\times+\frac{t}{2},t)
& =\frac{2\pi \sqrt{s}}{|\boldsymbol{p}|}   \sum_\ell (2\ell+1) d^\ell_{\lambda_1, -\lambda_3}(\theta(t,s)) \langle  \ell  \lambda_3  |\mathcal{M}^\dagger \mathcal{M} | \ell  \lambda_1 \rangle
\end{align}
where $\cos\theta(t,s)=1+2ts/(s-m^2)^2$. This exposes the fact that the arcs are integrals over the sum of positive-definite  {\it ``imaginary part''  matrices} $\mathcal{I}^{\ell}_{\lambda_1 \lambda_3}(s)$ modulated by some Wigner-$d$ function which is the only source of  controlled negativity, away from the forward limit  
\begin{align}
\label{arcPositiveDef1}
a^{(n)}_{\lambda_1 \lambda_3}(s,t)  = & \frac{2\sqrt{s}}{|\boldsymbol{p}|} \sum_{\ell\geq 2} (2\ell+1)\int_{s}^{\infty} \!\!\! \frac{ds^\prime}{(s^\prime-s_\times(t))^{2n+3}} d_{\lambda_1, -\lambda_3}^\ell(\theta(t,s^\prime)) \mathcal{I}^{\ell}_{\lambda_1 \lambda_3}(s^\prime) \,,\qquad n\geq 0\\ 
\label{arcPositiveDef2}
 \mathcal{I}^{\ell}_{\lambda_1 \lambda_3}(s^\prime) & =2\langle \ell  \lambda_1  |\mathcal{M}^\dagger \mathcal{M} | \ell  \lambda_3 \rangle   \succ  0\,. 
 \end{align}

  \subsubsection{The Positive Eikonal-Arcs}
  
So far we have been very general, with \eqref{arcPositiveDef1} and \eqref{arcPositiveDef2} holding even away from the eikonal limit. 
 Let's restrict now to the eikonal regime by projecting both sides of \eqref{arcPositiveDef1} with the Wigner-$d$ matrix and taking the limits $\ell\rightarrow \infty$, $|\boldsymbol{p}|\rightarrow \infty$.  
 
  
 Therefore, we act on both sides of \eqref{arcPositiveDef1}  with $\int_0^\infty dq^2 d^{\ell}_{\lambda_1 ,- \lambda_3}(\theta(q,s))$, after renaming $\sum_\ell \to \sum_{\ell^\prime}$.  As repeatedly done in the previous sections, we approximate the Wigner-$d$ in the large $\ell$ limit with a Bessel function \eqref{dlimitExplicit} (hence restricting to up to 2PM corrections, included). The integral is then dominated by the region at $\theta(q,s)\ll1$, but as $s^\prime>s$, this localizes as well the Wigner-$d$ appearing in \ref{arcPositiveDef1} around small values of $\theta(q,s^\prime)$. The small angle limit of the Wigner-$d$ deviates from the trivial result only for large $\ell$, implying that we can effectively replace also the second Wigner-$d$ by a Bessel function. We  
 replace the sum on $\ell^\prime$ with an integral on $b^\prime$, 
  obtaining \footnote{We used \eqref{eikonalLimit2} and \eqref{phasesinMbis}, and that $\sqrt{s} |\boldsymbol{p}|=(s-m^2)/2$.}
 \begin{equation}
 \label{pre-arc}
  \oint \frac{ds^\prime}{2\pi i } \frac{ \mathcal{M}_{\lambda_1 \lambda_3}(\boldsymbol{p},\mathbf{b})}{(s-m^2)^{2n+2}} =\frac{1}{\pi} \int_{s}^{\infty} \!\!\! \frac{ds^\prime}{(s^\prime-m^2)^{2n+2}} \int_0^\infty db^\prime b^\prime \int_0^\infty  dq q  J_{-\lambda_{3}-\lambda_1}(bq) J_{-\lambda_{3}-\lambda_1}(b^\prime q)      \widetilde{\mathcal{I}}_{\lambda_1 \lambda_3}(b^\prime, s^\prime)\, ,
 \end{equation}
where the two semi-circular contours are centered at $s^\prime=m^2$ and have radius $s-m^2$.  The matrix $\widetilde{\mathcal{I}}_{\lambda_1 \lambda_3}(s,b)\equiv c^2 e^{2i\lambda_{13}(\varphi -\pi/2)}\mathcal{I}^{\ell(b)}_{\lambda_1 \lambda_3}(s)\succ 0$ is positive definite because it differs from $\mathcal{I}^{\ell(b)}$ just by an unitary transformation and an irrelevant positive factor $c^2>0$.

Using the orthogonality condition \eqref{BesselOrthogonality} among Bessel functions we trivialize the integrals on the r.h.s. except for the one in $s^\prime$ which is over a manifestly positive-definite matrix, thus implying the positivity of the {\it ``eikonal arcs''} of finite radius: 
\begin{equation}
\label{eikonalArcs}
a^{(n)}_{\lambda_1 \lambda_3} (s,\mathbf{b})\equiv \oint \frac{ds^\prime}{2\pi i } \frac{\mathcal{M}_{\lambda_1 \lambda_3}(\boldsymbol{p}^\prime,\mathbf{b})}{(s^\prime-m^2)^{2n+2}}  \succ 0 
\end{equation}
where we remind that $\mathcal{M}_{\lambda_1 \lambda_3}(\boldsymbol{p},\mathbf{b})$ is the Eikonal transform \eqref{eikonalLimit3}. 
More explicitly, this means 
\begin{equation}
\label{nonlineareikarcs}
a^{(n)}_{+-} >0 \,, \qquad a^{(n)}_{-+} >0\,,\qquad a^{(n)}_{+-} a^{(n)}_{-+} - a^{(n)}_{++}a^{(n)}_{- -}  >0  \,, 
\end{equation}
  where we leave implicit the $b$ and $s$ dependence.  
  
In the limit where the scalar is very massive, it is convenient to change integration variable to $s^\prime =m^2+2m\omega$, with \eqref{eikonalArcs} becoming a statement about arcs in the graviton's frequency $\omega$, around the origin $\omega=0$ and radius $2m\omega$ 
  \begin{equation}
  \label{eikonalTimeDelArc}
 \oint \frac{d\omega}{2\pi i } \frac{\mathcal{M}_{\lambda_1 \lambda_3}(\omega,\mathbf{b})}{\omega^{2n+2}}  \succ 0.
  \end{equation}

These arcs are calculable in the EFT in terms of Wilson coefficients and therefore the  inequalities \eqref{nonlineareikarcs} and \eqref{eikonalTimeDelArc} provide positivity bounds on the EFT parameters.  Whenever the IR branch cuts can be neglected, these arcs are very simple to calculate, as they are given by the $(2n+1)$ $\omega$-derivative of the eikonal amplitude. 
For instance, for $n=0$ neglecting the IR branch cuts, we have thus proven the positivity of the time-delay matrix 
\begin{equation}
\label{posTimedelayMatrixProof}
\mathcal{T}_{\lambda_1 \lambda_3}(\omega,\mathbf{b})\equiv 
2\frac{\partial}{\partial\omega}\delta_{\lambda_1 \lambda_3}(s,\mathbf{{b}})\big|_{\omega=0} \succ 0  
\end{equation}
hence of its eigenvalues, since $\alpha_g\propto \omega$ while any higher derivative operator or any iteration of the eikonal exponentiation  can only increase the powers in $\omega$, and as such it can be discarded in the expression above that selects the value of the derivative at $\omega=0$.

From \eqref{eikonalTimeDelArc} it follows that higher values of $n$ constrain further higher odd $\omega$-derivatives of the eikonal amplitude 
  \begin{equation}
  \label{timeposw}
 \frac{\partial^{2n+1}}{\partial\omega^{2n+1}} \mathcal{M}_{\lambda_1 \lambda_3}(\omega,\mathbf{b})\big|_{\omega=0}\succ 0 \,, 
  \end{equation}
 neglecting the IR branch cuts.
 
  Including the IR branch cuts means working instead with the fully accurate arcs  \eqref{nonlineareikarcs} or \eqref{eikonalTimeDelArc} at finite radius. Similarly to the analysis in \cite{Bellazzini:2020cot}, the finite radius of the eikonal arcs regulates the IR soft divergences that appear to higher PM or gauge order, and that (partly) controls the running corrections to the Wilson coefficients. For example, the 2PM-1gauge corrections to the eikonal phase in the probe limit are expected to give rise to  $\delta=\alpha_g \left[-\log b/b_{\mathrm{IR}} + \frac{c}{2} \alpha (R_s/b) \left(\log\omega+\log -\omega \right)\right] $ with $c$ some number to be determined. Since $s$- or $\omega$-derivatives of arcs are negative definite, we anticipate that $c<0$.  These log corrections change the arc as $a^{(0)}(\omega, \mathbf{b})=2R_s\left(-\log b/b_{\mathrm{IR}}+ c \alpha (R_s/b) \log \omega\right)$, which is perfectly finite as long as $\omega$ is non-vanishing. 
  
 Notice, moreover,  that in pure gravity (no gauge corrections)  there is actually no IR log until 3PM. Furthermore,  to consistently go to 3PM and higher orders one should  extend \eref{pre-arc} by using the all-order expression \eqref{dUniformlimit} rather than its approximation \eqref{dlimitExplicit}.

The connection to higher even derivatives of the time-delay matrix  is no longer as direct as in \eqref{posTimedelayMatrixProof}, because \eqref{eikonalArcs} and \eqref{eikonalTimeDelArc} are given in terms of $i\mathcal{M}(\omega, \mathbf{b})=e^{2i\delta}-\mathbb{I}$ rather than in terms of $\delta$ itself~\footnote{Positivity of the even derivatives of time-delay matrix, as opposed to the odd-derivatives of $\mathcal{M}(s,\mathbf{b})$,  was used  in \cite{AccettulliHuber:2020oou} to argue for positivity bounds, but a derivation  of this claim was not presented. }.
For example, the $\omega^3$-term in $\mathcal{M}(\omega,\mathbf{b})$ can arise from a higher derivative operator or just from three iterations of $\delta_0$ in the eikonal exponentiation. 
Let's discuss this in an explicit example below. 
  
  \subsubsection{Example $R_{\mu\nu\rho\sigma}^4$}
  \label{subsec:ExampleRiemann4}
  
  An immediate application of the positivity of \eqref{timeposw} is on the higher derivative terms such as
  \begin{equation}
\mathcal{S}\supset \frac{1}{16\pi G} \int d^4x\sqrt{-g}\left[ -R+ \beta_1 (R_{\mu\nu\alpha\beta}R^{\mu\nu\alpha\beta})^2+ \beta_3 (R_{\mu\nu\alpha\beta}\epsilon^{\alpha\beta}_{\gamma\delta}R^{\gamma \delta\mu\nu})^2  \right]  \, .
  \end{equation}
  The contribution to the time delay has been computed in \cite{AccettulliHuber:2020oou} and is given by
  \begin{align}
\label{R4timed}
\Delta \delta_{\lambda_1 \lambda_3}(s,\boldsymbol{b})=\frac{315 \pi }{16} \frac{G^2 m^2\omega^3}{b}
\begin{pmatrix}
 \frac{\tilde\beta}{b^4}& \frac{\beta}{16 b_+^4}  \\
\frac{\beta}{16 b_-^4}   &   \frac{\tilde\beta}{b^4}
\end{pmatrix}
\end{align}
where $b_{\pm}=(b_1\pm i b_2)/2$, $\beta=4(\beta_1-\beta_3)$ and $\tilde\beta=4(\beta_1+\beta_3)$. Notice that this contribution to the time delay scales with $\omega^2$ compared to the linear dependence of the leading effect. 
However, the third iteration in the eikonal exponentiation of $\delta_0$ produces a contribution to $\mathcal{M}\sim (Gm \omega)^3  \log^3b/b_{\mathrm{IR}}$, and therefore the positivity of the first eigenvalue of $\partial^3\mathcal{M}(\omega=0,\mathbf{b})/\partial \omega^3$ translates into the condition  that $\tilde\beta$ can't be too negative, parametrically $\tilde\beta/b^6  \gtrsim -R_s/b$ (up to log's), where $b$ can be lowered up to reaching $b\sim 1/\Lambda$, i.e. the scale of new dynamics UV-completing the (Riemann)$^4$ operators.  In the limit $R_s \Lambda\to 0$, one thus get positivity of $\tilde\beta$.  Analogous arguments constrain $\beta$ not to be too negative, and demand positivity only in the limit $R_s \Lambda\to 0$, namely 
\begin{equation}
\label{eq:posbetas}
\beta \geq 0 \,\quad \tilde\beta \geq 0  \qquad \mbox{up to $O(\Lambda/\mpl \times m/\mpl)$ corrections.}
\end{equation}
That is, positivity of $\beta$ and $\tilde\beta$ is recovered in the ``weak and rigid limit'' $\Lambda/\mpl \times m/\mpl\to 0$, essentially  QFT on a fixed and nearly flat background for $m\to\infty$ and $\mpl\to\infty$, keeping fixed their ratio and $\Lambda$. These results are  in agreement with the positivity bounds  \cite{Bellazzini:2015cra} obtained away from the eikonal and rigid limit.

\subsection{Discussion on Causality}

We  have showed that ``asymptotic causality'' in a scalar-tensor theory, i.e. positive definiteness of the time-delay matrix for graviton-scalar scattering, is a direct consequence of analyticity and unitarity exploited in the eikonal regime.  Moreover, other positivity bounds on higher derivative EFT coefficients  follow as well in the  rigid limit of QFT coupled to gravity, see \eqref{eq:posbetas} for bounds on (Riemann)$^4$ operators. 
 
Since analyticity follows from micro-causality, see e.g  \cite{Weinberg:1995mt,Sommer:1970mr,Dubovsky:2007ac}, we have basically shown the implication  $\mathit{Micro-Causality}\Longrightarrow \mathit{Asymptotic\,\,\, Causality} $ 
in the context of gravitational physics. 
In other words, QFT can tolerate non-vanishing correlators at spacelike separation as long as the commutator of observables vanishes there, and this in turn implies that asymptotic causality ---positive time-delay--- must be respected.  This is a condition on the global causal structure of the theory,  extracted unambiguously from the eikonal amplitude. 

It does not imply, however, that signals should be confined inside some local, suitably-defined, light-cone.  For example, Ref.~\cite{Bellazzini:2021shn} showed that ``bulk causality'', the notion that all species experience longer time-delay than gravitons,  e.g. $\mathcal{T}_{\gamma} - \mathcal{T}_{g} \geq 0$, is actually violated quantum mechanically by a resolvable amount.   Likewise, ``infrared causality''  demanding positivity w.r.t. Shapiro time delay $\mathcal{T}_{\gamma} - \mathcal{T}_{\mathcal{\rm Sh}} \geq 0$ \cite{Chen:2021bvg},  is violated as well quantum mechanically in full QED with dynamical charged states of mass $m_e$,  by a resolvable amount as soon as $b \ll 1/m_e$. In fact, infrared and bulk causality are essentially the same condition since  $\mathcal{T}_g \simeq \mathcal{T}_{\rm Sh}$ within the 1PM approximation.  Notice that for $b\ll 1/m_e$, removing the EFT=QED contribution  to $\mathcal{T}$ as prescribed in \cite{Chen:2021bvg}  corresponds to push to infinity the Landau pole by either trivializing the gauge theory with a vanishing gauge coupling, or to go outside this regime with $m_e\to\infty$, where IR and bulk causality are again the same. 

  In summary, propagation outside a local light-cone defined by the graviton can take place within a gauge theory coupled to gravity, as long as the propagation is still confined within the asymptotic Minkowski light-cone giving $\mathcal{T}\geq 0$. 

 As shown in the previous (Riemann)$^4$-example in \ref{subsec:ExampleRiemann4},  positivity bounds on higher odd derivatives of the eikonal amplitude \eqref{timeposw} allow us to target contact terms by inserting them into loops for the scattering phase matrix.   One could extract even sharper bounds by working away from the eikonal and rigid limit using \eqref{arcPositiveDef1} along the lines of \cite{Bellazzini:2021oaj,Caron-Huot:2021rmr,Caron-Huot:2022ugt,Henriksson:2022oeu}. We leave this task to future work.

\section{Conclusions}
\label{sec:concl}

In this work we have investigated the eikonal scattering of two bodies interacting via gravity. The eikonal amplitude is obtained by studying the limit of large angular momentum of the system expanded in partial waves. 
We have shown the geometrical picture behind the eikonal expansion is the same of the flat-earth approximation, where the  $SU(2)$ group of rotations contracts to $ISO(2)$. Moreover,  the resulting continuous-spin representation allows us to recover the 2D Fourier transform to $b$ space, as well as the classical limit of continuous angular momentum. 

While physically intuitive and compelling, we have shown that this approach is valid  up to $\theta^2$ corrections.  We have thus extended these results by finding an expression for the eikonal amplitude valid at all orders in $\theta$, extending in principle the validity of approximation beyond the small angle regime. What is the radius of convergence of this expansion? Can we use the all-order expression to approach the region of black hole formation?   These remain, at the moment, open and interesting questions.

We have also shown that quantum gravity corrections are never resolvable within the eikonal transplanckian regime, i.e. they are always smaller than the quantum fuzziness on the trajectories.    Moreover, we have explained under what conditions classical corrections are not always the largest ones, and we have provided an example of a gauge theory coupled to gravity where  quantum corrections to the eikonal scattering angle that are more important than the classical Post-Minkowskian's.  
We have argued that ``classical'' as opposed to ``quantum'' is not quite an useful distinction for  the various source of the eikonal corrections. What matters instead in an EFT is just the largest length  scale ---provided it has a non-negligible coupling associated--- regardless of its classical or quantum origin, that a low-energy observer cannot possibly know anyway. 

In this respect, we have studied QFT contributions to eikonal scattering in a gauge theory coupled to gravity, and contrasted  it with the Post-Minkowskian corrections, working to two-loop order.  We have explored the $O(G^2 \alpha)$ term in the scattering of photons off some massless neutral scalar field, where $\alpha$ is the fine structure constant.  
As in the purely massless gravitational case, however, we have found no resolvable effect beyond the $O(G \alpha)$ contribution to the eikonal that has been originally calculated in \cite{Bellazzini:2021shn}, and which is nevertheless larger than pure to 2PM corrections as soon as $(R_s/b)^2\ll \alpha$. 

We plan to investigate further this problem in future work by  introducing  a mass to the scalar spectator and exploit the same technology to compute the expected 2PM$+$1Gauge resolvable  effect. 
 In particular, in the probe limit for the photons, the differential equations relating Master Integrals are still a function of one kinematic variable, as the dependence on the mass of the scalar trivializes. This means that most of  techniques used in this paper can be directly exported  into the probe scenario.

Finally, we have investigated the causal properties of the eikonal amplitude including the QFT corrections. 
By introducing certain  ``eikonal arcs'' in complex energy space at fix impact parameters, we have shown that the time delay and odd higher derivatives of the eikonal amplitude  satisfy infinitely many non-linear constraints, that can be used to bound EFT coefficients. In an explicit example of this strategy, we have reproduced the known positivity bounds for (Riemann)$^4/\mpl^2\Lambda^4$ type of operators, in the rigid limit $\Lambda/\mpl\times m/\mpl\to 0$, where $m$ is the mass of one of the scattered particles.  It is quite remarkable that eikonal scattering allows to assess positivity bounds for contact operators by including them into the loop corrections to the scattering phase.  

A possible future direction is to explore these new causality  bounds more systematically, away from eikonal and rigid limits, using \eqref{arcPositiveDef1}.  In particular, 5-point contact terms would appear in the time delay as two loops contributions, and can in principle be targeted by our bounds,  circumventing the complications of knowing the analytic structure of 5-point correlation functions. Ultimately, we expect that $2$-to-$n$ processes should be constrained as well by fundamental principles.

\subsection*{Acknowledgments}
We thank Tim Anson,  Fabian Bautista, Stefano De Angelis, Kelian H\"aring, Johannes Henn, Philipp Kreer,  Ben Page, Rodolfo Russo, Javi Serra, Francesco Sgarlata, Ofri Telem, Pierre Vanhove, Filippo Vernizzi, and Sasha Zhiboedov  for useful discussions. 
M.M.R. is funded by the Deutsche Forschungsgemeinschaft (DFG, German Research Foundation) under Germany's Excellence Strategy -- EXC 2121 ``Quantum Universe'' -- 390833306. The work of G.I.  is supported by the Swiss National Science Foundation under grant no. 200021\_205016.

\appendix

\section{The IR-divergent Coulomb Phase via Wordline Eikonal}
\label{wordline-Eik}

In this short appendix we discuss one particular type of infrared (IR) divergence that arises already in the leading eikonal amplitude: the divergent Coulomb phase. We would like to provide a simple physical understanding of this IR effect. 

 Given a scattering amplitude, one can define an associated potential by matching to the amplitude that it would produce,   
\begin{equation}
V(\boldsymbol{x})= N \int \frac{d^3 \boldsymbol{q}}{(2\pi)^3} e^{-i \boldsymbol{q} \boldsymbol{x}} \mathcal{M}(s, \boldsymbol{q})\,, 
\end{equation}
up to a normalization factor $N$ that we don't need in the following. Inverting this relation and expressing $\mathcal{M}$ in the eikonal  limit \eqref{eikonalLimit4} one gets 
\begin{equation}
\label{worline-EikEq1}
e^{2i\delta(s,\mathbf{b})}-\mathbb{I}=\frac{i}{4|\boldsymbol{p}| \sqrt{s} N } \int_{-\infty}^{\infty} \!\!\!dz   V(\boldsymbol{x})\big|_{\boldsymbol{x}=(z,\mathbf{b})}  
\end{equation}
which reproduces the textbook result in potential scattering theory \cite{landau3}. 
  The interpretation of this result is straightforward now: the leading eikonal amplitude in impact parameter space is just the wordline integral of the potential over the straightline geodesic. 
  
This formula  allows an immediate interpretation of the IR divergence for the time-delay coming from the $1/|\boldsymbol{x}|=1/\sqrt{z^2+\mathbf{b}^2}$-potential, which itself originates from the 3D Fourier-transform of amplitudes with a $1/\boldsymbol{q}^2$-pole of the massless graviton exchanged in $t$-channel.  At the lowest  order in $G$ and integrating over a finite travelled distance $L$ from the source to the detector, 
\begin{equation}
\delta(s,\mathbf{b}) \propto - \frac{1}{4|\boldsymbol{p}| \sqrt{s} N } \log b/L+\ldots 
\end{equation} 
The IR divergence arises because the time-delay is accumulated ---logarithmically--- over the travelled distance $L$. Since the source and the detector are always at some finite distance, we can choose $L$ as small as the largest length scale we want to include in the scattering.  For instance, if we want to probe heavy physics at the mass scale $\Lambda$ (no new massless modes), then it's enough to choose $L$ a couple of orders of magnitude larger than $1/\Lambda$. The precise hierarchy is not crucial, because it would impact the accuracy of the results only logarithmically. 
A similar IR divergence arises in QED but in that case some screening mechanism can always be devised by adding spectator charges to neutralise the system as seen at long distance. 

Other IR divergences than Coulomb's can arise to higher orders. 
In a purely gravitational theory there are soft divergences only \cite{Weinberg:1965nx,Akhoury:2011kq} , which cancel once including the emission of real gravitational radiation, e.g. using the eikonal-operator approach \cite{Ciafaloni:2018uwe,DiVecchia:2022piu}.  On the other hand, when also light matter fields are present, collinear divergences can be present too. For example, \cite{Bellazzini:2021shn} has shown how to deal with Sudakov (soft and collinear)  IR divergences in the context of gravitational eikonal scattering at 1PM-1gauge order.   A general treatment is however lacking at the moment, and it represents one of the limitations to export sharp and IR-safe positivity bounds in the context of gravity.

\section{Partial Waves: full monty}
\label{AppPhases}

In this appendix we review how to obtain from scratch  the partial wave expansion for (distinct) particles of any mass and spin, reproducing the classic results of \cite{Jacob:1959at}. 
The extension to identical particles is straightforward.  

Let's consider irrep of the Poincar\'e group which are labelled by the 4-momentum $p$ with $p^0> 0$, by the total angular $\boldsymbol{J}^2=\ell(\ell+1)$ and its projection $\lambda$ (the helicity) along the direction of motion, and possibly by some other internal collective label $\alpha=\{q, m, \lambda_i \ldots \}$ such as the conserved charges, the particle type and its mass, the helicities $\lambda_i$ of the constituents etc, that is needed to specify the irreps uniquely: 
\begin{equation}
\label{scalarprod3}
| p \,\ell\, \lambda, \alpha\rangle\,,  \qquad  \langle p^\prime\, \ell^\prime\, \lambda^\prime ,\alpha^\prime |p\, \lambda\, \ell , \alpha \rangle=  (2\pi)^4 \delta^4 (p-p^{\prime}) \delta_{\ell \ell^\prime} \delta_{\lambda \lambda^\prime}\delta_{\alpha\alpha^\prime}\,.
\end{equation}
We have chosen the relativistic normalization and $\delta_{\alpha\alpha^\prime}$ represents the  product of Kronecker deltas $\delta_{qq^\prime}\delta_{mm^\prime}\delta_{\lambda_i^\prime \lambda_i}\ldots $ for the internal labels. 
   The resolution of the identity 
\begin{align}
\label{resId1}
&\mathbb{I}=   \sum_\alpha \sum_\ell \sum_{\lambda=-\ell}^{\ell}  \int \frac{d^4 p}{(2\pi)^4} \theta(p^0)\theta(p^2) | p\, \ell\, \lambda ;\alpha \rangle \langle p\, \ell\, \lambda ; \alpha |  \,,
\end{align}
allows to reconstruct any state from its projection on the irreps. 

Single particle states are examples of  irreps of definite mass $p^2=m^2$ (hence often labelled by 3-momentum and the mass $m$) where $\ell=S$ is their spin. 
The tensor product of two-particle states, represented by 
\begin{equation}
\label{def2particlestate}
| p_1\, S_1\, \lambda_1, \alpha_1\rangle  | p_2\, S_2\, \lambda_2, \alpha_2 \rangle  \equiv | p_1\, S_1\, \lambda_1, \alpha_1 ; p_2\, S_2\, \lambda_2, \alpha_2 \rangle\,,
\end{equation}
 has definite 4-momentum $p_1+p_2$ but it is in general reducible into the sum of irreps of various angular momenta, called the partial waves. 
 
 In order to determine the decomposition of a two-particle state in irreps, let's consider first the case of particles $i$ and $j$ moving along the $z$-axis in their c.o.m. frame,   $p_i=(\omega_i,0,0,p^z_{i}>0)$ and $p_j=(\omega_j,0,0,-p^z_i)$: by invariance under translations and rotations around the $z$-axis, the state has definite momentum $p_i+p_j=({\omega_i+\omega_j,\mathbf{0}})$ and definite helicity $\lambda_i-\lambda_j$, therefore the projection over an irrep is actually fixed
 \begin{equation}
 \label{projectionzaxis}
\langle  p \,\ell\, \lambda, \alpha | p_i\, S_i\, \lambda_i,\alpha_i ; p_j\, S_j\, \lambda_j, \alpha_j \rangle_{\substack{ \\ c.o.m. \\  \!\theta=\phi=0}}=(2\pi)^4\delta^4(p-\bar{p}_{ij})\delta_{\lambda \lambda_{ij}}\delta_{\alpha\bar{\alpha}_{ij}} C_\ell(\boldsymbol{p}_{i}^{2},\alpha)
 \end{equation}
 up to some  weight factor $C_\ell(\boldsymbol{p}_{i}^{2},\alpha)$ that is going to be fixed by a normalization condition, and where we introduced the notation
  \begin{equation}
 \bar{p}_{i j}=p_i+ p_j \,,\qquad \lambda_{ij}=\lambda_i - \lambda_j\,,\qquad \bar{\alpha}_{ij}=\alpha_i\cup \alpha_j=\{q_i, q_j,m_i, m_j, \lambda_i, \lambda_j ,\ldots \} \,.
 \end{equation}
 We also used that all the invariant scalar products are just functions of the c.o.m. tree-momentum squared $|p_i^{z}|=\boldsymbol{p}_{i}^2=\boldsymbol{p}_{c.\,i}^{2}=\boldsymbol{p}_{c.\, j}^{2}$, which can be written in a covariant form as 
 \begin{equation}
 \label{com3momentum}
 \boldsymbol{p}_{c.\, i}^{2}=\boldsymbol{p}_{c.\, j}^{2}=\frac{1}{4s_{ij}} \left(s_{ij}-(m_i+m_j)^2\right)\left((s_{ij}-(m_i-m_j)^2\right))   \qquad s_{ij}=(p_i+p_j)^2\,.
 \end{equation}
 The overlap \eqref{projectionzaxis} corresponds, via \eqref{resId1}, to the decomposition over infinite partial waves\footnote{For ease of notation, whenever the 3-momentum of the particles is the same as the c.o.m. 3-momentum we omit the subscript $c.$ unless needed otherwise.}
 \begin{equation}
 \label{1stdecomposition}
 | p_i\, S_i\, \lambda_i,\alpha_i ; p_j\, S_j\, \lambda_j, \alpha_j \rangle_{\substack{ \\ c.o.m. \\  \!\theta=\phi=0}}= \sum_{\ell\geq |\lambda_{ij}|} C_\ell(\boldsymbol{p}_{i}^{2},\bar{\alpha}_{ij}) \,  | \bar{p}_{ij}\,\ell\, \lambda_{ij}, \bar{\alpha}_{ij} \rangle 
 \end{equation}
From this expression we can now obtain the irrep decomposition for a generic state by suitable boosts and rotations. For example, boosting \eqref{1stdecomposition} along the z-direction we get 
\begin{equation}
 \label{2nddecomposition}
 | p_i\, S_i\, \lambda_i,\alpha_i ; p_j\, S_j\, \lambda_j, \alpha_j \rangle_{\substack{ \\  \!\theta=\phi=0}}= \sum_{\ell\geq |\lambda_{ij}|} C_\ell(\boldsymbol{p}_{c.\, i}^{2},\bar{\alpha}_{ij}) \,  | \bar{p}_{ij}\,\ell\, \lambda_{ij}, \bar{\alpha}_{ij} \rangle \,. 
 \end{equation}

Or, to get from the $\theta=\phi=0$ configuration to a generic one (but still in the c.o.m.), we can apply a rotation $R(\phi,\theta,-\phi)$  where we define $R(\alpha,\beta,\gamma)=e^{-i\alpha J^3}e^{-i\beta J^2}e^{-i\gamma J^3}$. Observing that in the c.o.m. frame $\bar{p}_{ij}=R\bar{p}_{ij}$  behaves just like any other internal index transparent to rotations,  and reminding 
the definition of the Wigner-$d$ matrix (oblivious of trivial internal indices, and chosen real) 
\begin{equation}
\label{WignerDDef}
 \langle  \ell \lambda |e^{-i\theta J^2} | \lambda^\prime \ell^\prime \rangle \equiv \delta_{\ell\ell^\prime} d^{\ell}_{\lambda \lambda^\prime }(\theta)
\end{equation}
we get  
\begin{equation}
R(\phi,\theta,-\phi)  | \bar{p}_{ij}\,\ell\, \lambda_{ij}, \bar{\alpha}_{ij} \rangle_{\substack{ \\ c.o.m. \\  \!\theta=\phi=0}}=\sum_{|\lambda|\leq \ell} e^{-i(\lambda-\lambda_{ij})\phi}d^{\ell}_{\lambda \lambda_{ij}}(\theta)\,  | \bar{p}_{ij}\,\ell\, \lambda, \bar{\alpha}_{ij} \rangle
\end{equation}
which in turn gives 
\begin{equation}
 \label{3rddecomposition}
R(\phi,\theta,-\phi) | p_i\, S_i\, \lambda_i,\alpha_i ; p_j\, S_j\, \lambda_j, \alpha_j \rangle_{\substack{ \\ c.o.m. \\  \!\theta=\phi=0}}= \sum_{\ell\,\lambda}
 C_\ell(\boldsymbol{p}_{i}^{2},\bar{\alpha}_{ij}) e^{-i(\lambda-\lambda_{ij})\phi}d^{\ell}_{\lambda \lambda_{ij}}(\theta) \,  | \bar{p}_{ij}\,\ell\, \lambda, \bar{\alpha}_{ij} \rangle \,.
 \end{equation}
The summation domain $\ell\geq |\lambda_{ij}|\,,\, |\lambda|\leq \ell$ is left understood hereafter.  Notice, however, that the left-hand side of \eqref{3rddecomposition} is the tensor product of the rotated one-particle states only up to an overall $e^{-2i\lambda_j \phi}$-phase.   Therefore, the decomposition would actually read 
 \begin{equation}
  \label{4thdecomposition}
e^{2i\lambda_j \phi} | p_i\, S_i\, \lambda_i,\alpha_i ; p_j\, S_j\, \lambda_j, \alpha_j \rangle_{\substack{ \\ c.o.m.}}=\sum_{\ell\,\lambda} C_\ell(\boldsymbol{p}_{i}^{2},\bar{\alpha}_{ij}) e^{-i(\lambda-\lambda_{ij})\phi}d^{\ell}_{\lambda \lambda_{ij}}(\theta) \, | \bar{p}_{ij}\,\ell\, \lambda, \bar{\alpha}_{ij} \rangle \,.
  \end{equation}
  However, as it is customary in the literature since \cite{Jacob:1959at},  we actually absorbe that $e^{2i\lambda_j \phi}$-factor  in the  {\it definition of the two-particle scattering state}, see \eqref{statesdef1} and \eqref{statesdef2}. 
  
The first of the $d$-matrix orthogonality conditions
  \begin{align}
  \label{orth1}
 \int_{-1}^1 \!\!\!\!\!\!d\!\cos\theta \,\,d^{\ell^\prime}_{\lambda^\prime \lambda}(\theta) d^{\ell}_{\lambda^\prime \lambda}(\theta) = \frac{2}{2\ell+1}\delta_{\ell^\prime \ell}   \,,
 \qquad 
\sum_\ell \frac{(2\ell+1)}{2}d^\ell_{\lambda^\prime \lambda}(\theta) d^\ell_{\lambda^\prime \lambda}(\theta^\prime) =\delta(\cos\theta-\cos\theta^\prime)
 \end{align}
allow to invert the relation between the 2-particle states and the irreps 
\begin{equation}
\label{inverseformula}
| \bar{p}_{ij}\,\ell\, \lambda, \bar{\alpha}_{ij} \rangle=\frac{(2\ell+1)/2}{C_\ell(\boldsymbol{p}_{i}^{2},\bar{\alpha}_{ij})}\int_{-1}^1 \!\!\!\!\!\!d\!\cos\theta \!\! \int_{0}^{2\pi} \! \frac{d\phi}{2\pi} 
e^{-i(\lambda_{ij}-\lambda)\phi} d^{\ell}_{\lambda \lambda_{ij}}(\theta) \left(e^{i2\lambda_j \phi} | p_i\, S_i\, \lambda_i,\alpha_i ; p_j\, S_j\, \lambda_j, \alpha_j \rangle_{\substack{ \\ c.o.m.}}\right)
\end{equation}
where the left-hand side ``knows'' about the $\lambda_i$ via its internal parameter $\bar{\alpha}_{ij} $. 

Finally, we can determine the weight factor $C_\ell(\boldsymbol{p}_{i}^{2},\alpha)$ by matching the normalization induced by the  one-particle tensor product, that  in spherical coordinates reads 
 \begin{align}
\label{norm1change}
\langle p_3\, S_3\, \lambda_3, \alpha_3 ;  p_4\, S_4\, \lambda_4, \alpha_4 & |p_1\, S_1\, \lambda_1, \alpha_1 ; p_2\, S_2\, \lambda_2,\alpha_2 \rangle_{\substack{ \\  \rm c.o.m.}}   =\\
 \nonumber
&= (2\pi)^4  \delta^4(\bar{p}_{12}-\bar{p}_{34}) 16\pi^2 \sqrt{\frac{s}{|\boldsymbol{p}_{1}| |\boldsymbol{p}_{3}|}} \delta(\cos\theta-\cos\theta^\prime)\delta(\phi-\phi^\prime)\delta_{\lambda_1 \lambda_3} \delta_{\lambda_2\lambda_4}\delta_{\bar{\alpha}_{12}\bar{\alpha}_{34}}\,, 
\end{align}
with the normalization implied by \eqref{scalarprod3} for the irrep on the left-hand side of \eqref{inverseformula}, yielding 
\begin{equation}
\left|C_{\ell}(\boldsymbol{p}_{i}^{2}, \bar{\alpha}_{ij})\right|=  4\pi(2\ell+1)\frac{\sqrt{s_{ij}}}{|\boldsymbol{p}_{i}|}\,. 
\end{equation}
This is  determined only up to a phase that we reabsorb in the definition of the states. Moreover, it is actually independent on $\bar{\alpha}_{ij}$ (except for the implicit dependence on $m_i$ in $s_{ij}$) which will no longer be displayed.  

\subsection*{Miscellaneous identities}

\begin{equation}
\label{DinJacobi}
d^\ell_{\lambda^\prime \lambda}(\theta) =(-1)^{\lambda^\prime-\lambda}\sqrt{\frac{(\ell+\lambda^\prime)! (\ell-\lambda^\prime)!}{(\ell+\lambda)! (\ell-\lambda)!}}\left(\sin\frac{\theta}{2}\right)^{\lambda^\prime-\lambda} \left(\cos\frac{\theta}{2}\right)^{\lambda+\lambda^\prime}  P^{(\lambda^\prime-\lambda,\lambda+\lambda^\prime)}_{\ell-\lambda^\prime}(\cos\theta)
\end{equation} 
for $\lambda^\prime \geq \lambda$; whereas  for $\lambda^\prime <\lambda$ one can use $d^\ell_{\lambda^\prime \lambda}(\theta)=(-1)^{\lambda^\prime-\lambda}d^{\ell}_{\lambda \lambda^\prime}$. The $P^{(a,b)}_{\ell}(\cos\theta)$ are Jacobi polynomials in $\cos\theta$. 
\begin{equation}
\label{BesselOrthogonality}
\mbox{Bessel Orthogonality:}\quad  \int_0^\infty dq q J_\nu (q a) J_\nu (q b)=\delta(a-b)/a
 \end{equation}

\section{Two-loop master integrals}
\label{app:MIs}

In this brief appendix we shall give the explicit expression of the two-loop scalar MIs sketched in \fref{fig:MIs}, \sref{sec:2PMapl}. Let us start with the planar MIs. Using the external momenta defined in \eref{eq:pbardef} and calling the loop momenta $\ell_1$ and $\ell_2$, we introduce the following propagators
\begin{equation}
\begin{aligned}
	\rho_1 & = \ell_1^2 \, , & \rho_2 & = \left(\ell_1 + \bar{p}_2-\frac{q}{2}\right)^2 \, , & \rho_3 & = \left(\ell_1 + \bar{p}_1 +\bar{p}_2\right)^2 \, , \\
	\rho_4 & = \left(\ell_1+\bar{p}_2+\frac{q}{2}\right)^2 \, , & \rho_5 & = \ell_2^2\, , & \rho_6 & = \left(\ell_2 + \bar{p}_2-\frac{q}{2}\right)^2 \, , \\
	\rho_7 & = \left(\ell_1 + \bar{p}_1 +\bar{p}_2\right)^2 \, , & \rho_8 & = \left(\ell_2+\bar{p}_2+\frac{q}{2}\right)^2\, , & \rho_9 & = \left(\ell_1-\ell_2\right)^2\, ,
\end{aligned}
\end{equation}
where we left implicit the $+i 0^+$ Feynman prescription. Then, all the MIs take the following form
\begin{equation}
	G_{\underline{i_1},i_2,i_3,i_4,i_5,i_6,\underline{i_7},i_8,\underline{i_9}} \equiv \int \dint{d}{\ell_1} \dint{d}{\ell_2} \frac{1}{\underline{\rho}_1^{i_1} \rho_2^{i_2} \rho_3^{i_3} \rho_4^{i_4} \rho_5^{i_5} \rho_6^{i_6} \underline{\rho}_7^{i_7} \rho_8^{i_8} \underline{\rho_9}^{i_9}} \, .
\end{equation}
In the above equation we underlined the propagators that are cut as depicted in \fref{fig:MIs}. These propagators are to be replaced with on-shell delta functions \cite{Herrmann:2021tct,DiVecchia:2021bdo,Riva:2021vnj}, e.g
\begin{equation}
	\frac{1}{\underline{\rho}_7} \to \dd_+(\rho_7) \, .
\end{equation} 
Higher power of a cut propagators means derivative of the delta functions. Given these definitions, the planar MIs $\vec{g}$ are explicitly
\begin{equation}
	\begin{aligned}
		g_1 & = \frac{1}{(-q^2)}\frac{ \left(-18 \varepsilon ^3+27 \varepsilon ^2-13 \varepsilon +2\right)}{4 (y +1) \varepsilon ^3}G_{\underline{1}, 0, 0, 0, 0, 0, \underline{1}, 0, \underline{1}} \, , \\
		g_2 & = (-q^2)\left(1-\frac{1}{2\varepsilon}\right)G_{\underline{1}, 0, 0, 0, 0, 1, \underline{1}, 1, \underline{1}} \, , \\
		g_3 & = (-q^2)\frac{1-y}{2}G_{\underline{1}, 0, 0, 1, 0, 1, \underline{1}, 0, \underline{1}} \, , \\
		g_4 & = -(-q^2)^3\frac{1+y}{4}G_{\underline{1}, 1, 0, 1, 0, 1, \underline{1}, 1, \underline{1}} \, , \\
		g_5 & = -\frac{72 g_1}{y +1}-\frac{18 g_2 \left(\varepsilon^2+5 \varepsilon +2\right)}{(y +1) \left(2 \varepsilon^2+3 \varepsilon +1\right)}+\frac{18 g_3}{y +1}-\frac{3 g_4 \left(4 y +\frac{1}{\varepsilon }+7\right)}{y +1} \\
		& \qquad \qquad -(-q^2)^4\frac{3 (y +1) (\varepsilon +1)}{8 \left(2 \varepsilon ^2+\varepsilon \right)} G_{\underline{2}, 1, 0, 1, 0, 1, \underline{1}, 1, \underline{1}}\, .
	\end{aligned}
\end{equation}
This planar MIs satisfy the differential equation written in \eref{eq:diffplanar} where $A_{\pm 1}$ are explicitly
\begin{align}
A_{+1} = \left(
\begin{array}{ccccc}
 -2 & 0 & 0 & 0 & 0 \\
 -4 & -2 & 0 & 0 & 0 \\
 -8 & 0 & -2 & 0 & 0 \\
 -24 & -12 & 6 & -2 & 0 \\
 396 & 180 & -99 & 39 & 1 \\
\end{array}
\right) \, , & &  A_{-1}= \left(
\begin{array}{ccccc}
 0 & 0 & 0 & 0 & 0 \\
 4 & 1 & 0 & 0 & 0 \\
 0 & 0 & 2 & 0 & 0 \\
 24 & 12 & -6 & 10 & \frac{2}{3} \\
 -396 & -180 & 99 & -135 & -9 \\
 \end{array}
  \right) \, ,
\end{align}

For the non planar MIs, we introduce the following basis of propagators
\begin{equation}
\begin{aligned}
	\tilde{\rho}_1 & = \ell_1^2 \, , & \tilde{\rho}_2 & = \left(\ell_1 + \bar{p}_2-\frac{q}{2}\right)^2 \, , & \tilde{\rho}_3 & = \left(\ell_1 - q \right)^2 \, , \\
	\tilde{\rho}_4 & = \ell_2^2 \, , & \tilde{\rho}_5 & = \left(\ell_2+\bar{p}_2-\frac{q}{2}\right)^2 \, , & \tilde{\rho}_6 & = \left(\ell_2 - q\right)^2 \, , \\
	\tilde{\rho}_7 & = \left(\ell_1 - \bar{p}_1 -\frac{q}{2}\right)^2 \, , & \tilde{\rho}_8 & = \left(\ell_1-\ell_2\right)^2 \, , & \tilde{\rho}_9 & = \left( \ell_1 -\ell_2 + \bar{p}_1 +\frac{q}{2}\right) \, ,
\end{aligned}
\end{equation}
and the definition
\begin{equation}
	\tilde{G}_{i_1,\underline{i_2},i_3,i_4,i_5,i_6,\underline{i_7},\underline{i_8}, i_9} \equiv \int \dint{d}{\ell_1} \dint{d}{\ell_2} \frac{1}{\rho_1 \underline{\tilde{\rho}}_1^{i_2} \tilde{\rho}_3^{i_3} \tilde{\rho}_4^{i_4} \tilde{\rho}_5^{i_5} \tilde{\rho}_6^{i_6} \underline{\tilde{\rho}}_7^{i_7} \underline{\tilde{\rho}_8}^{i_8} \rho_9} \, .
\end{equation}
The MIs $\vec{\tilde{g}}$ are then given by
\begingroup
\allowdisplaybreaks
	\begin{align}
		\tilde{g}_1 & = \frac{2-2\varepsilon}{3\varepsilon} g_1 \, , \qquad \tilde{g}_2 = \frac{2-2\varepsilon}{3\varepsilon} g_2 \, ,\qquad \tilde{g}_3 = \frac{2-2\varepsilon}{3\varepsilon} g_3 \, , \qquad 
		\tilde{g}_4 = (-q^2)\frac{2-2\varepsilon}{3\varepsilon} \tilde{G}_{0,\underline{1},0,0,0,1,\underline{1},\underline{1}, 1} \, , \notag \\
		\tilde{g}_5 & = \bigg[ \frac{y  \left(19346 \varepsilon ^4+26331 \varepsilon ^3+11948 \varepsilon ^2+2229 \varepsilon +146\right)-150 \varepsilon  \left(90 \varepsilon ^3+111 \varepsilon ^2+42 \varepsilon +5\right)}{25 y  \left(y ^2-1\right) (\varepsilon +1) (2 \varepsilon +1)^2 (4 \varepsilon +1)} \notag \\
& \quad +\frac{y  \left(y  \left(334 \varepsilon ^4-21 \varepsilon ^3-908 \varepsilon ^2-699 \varepsilon -146\right)-60 \varepsilon  \left(127 \varepsilon ^3+185 \varepsilon ^2+85 \varepsilon +13\right)\right)}{25 \left(y ^2-1\right) (\varepsilon +1) (2 \varepsilon +1)^2 (4 \varepsilon +1)}\bigg] \tilde{g}_1 \notag \\
& \quad +\frac{4 \left(7 y ^2 \left(43 \varepsilon ^2-69 \varepsilon -34\right)-y  \left(5999 \varepsilon ^2+3093 \varepsilon +238\right)+450 \varepsilon  (3 \varepsilon +1)\right)}{75 y  (y +1) \left(8 \varepsilon ^2+6 \varepsilon +1\right)} \tilde{g}_2 \notag \\
& \quad + \frac{y ^2 \left(-334 \varepsilon ^2+522 \varepsilon +292\right)+y  \left(5906 \varepsilon ^2+3282 \varepsilon +292\right)-300 \varepsilon  (3 \varepsilon +1)}{25 y  (y +1) \left(8 \varepsilon ^2+6 \varepsilon +1\right)}\tilde{g}_3 \notag \\
& \quad + \frac{48 \varepsilon  \left(5 y ^2 (\varepsilon +1)+7 y  \varepsilon -5 (3 \varepsilon +1)\right)}{5 y  \left(y ^2-1\right) (\varepsilon +1) (4 \varepsilon +1)} \tilde{g}_4 \notag \\
& \quad -(-q^2)^3\frac{2 (\varepsilon -1) \left(5 y ^2 (4 \varepsilon +1)-7 y  (3 \varepsilon +1)+10 \varepsilon +5\right)}{15 y  \varepsilon  (4 \varepsilon +1)} \tilde{G}_{1,\underline{1},1,0,0,1,\underline{1},\underline{1}, 1} \notag \\
& \quad +(-q^4)(\frac{(7 y -10) \left(y ^2-1\right) (\varepsilon^2 -1)}{30 y  \varepsilon  \left(8 \varepsilon ^2+6 \varepsilon +1\right)} \tilde{G}_{1,\underline{1},2,0,0,1,\underline{1},\underline{1}, 1} \, ,  \\
\tilde{g}_6 & = \bigg[ -\frac{8 \left(y  \left(26902 \varepsilon ^4+36747 \varepsilon ^3+16726 \varepsilon ^2+3123 \varepsilon +202\right)-225 \varepsilon  \left(90 \varepsilon ^3+111 \varepsilon ^2+42 \varepsilon +5\right)\right)}{25 y  \left(y ^2-1\right) (\varepsilon +1) (2 \varepsilon +1)^2 (4 \varepsilon +1)} \notag \\
&\quad + \frac{8 y ^2 \left(-229 \varepsilon ^2+357 \varepsilon +202\right) (2 \varepsilon +1)+360y  \varepsilon  \left(222 \varepsilon ^3+325 \varepsilon ^2+150 \varepsilon +23\right)}{25 \left(y ^2-1\right) (2 \varepsilon +1)^2 (4 \varepsilon +1)} \bigg] \tilde{g}_1 \notag \\
& \quad -\frac{12 \left(y ^2 \left(43 \varepsilon ^2-69 \varepsilon -34\right)-2 y  \left(466 \varepsilon ^2+237 \varepsilon +17\right)+75 \varepsilon  (3 \varepsilon +1)\right)}{25 y  (y +1) \left(8 \varepsilon ^2+6 \varepsilon +1\right)} \tilde{g}_2 \notag \\
& \quad + \frac{y ^2 \left(458 \varepsilon ^2-714 \varepsilon -404\right)-4 y  \left(2068 \varepsilon ^2+1146 \varepsilon +101\right)+450 \varepsilon  (3 \varepsilon +1)}{25 y  (y +1) \left(8 \varepsilon ^2+6 \varepsilon +1\right)} \tilde{g}_3 \notag \\
& \quad -\frac{72 \varepsilon  \left(5 y ^2 (\varepsilon +1)+6 y  \varepsilon -5 (3 \varepsilon +1)\right)}{5 y  \left(y ^2-1\right) (\varepsilon +1) (4 \varepsilon +1)} \tilde{g}_4 \notag \\
& \quad + \frac{(\varepsilon -1) \left(5 y ^2 (4 \varepsilon +1)-6 y  (3 \varepsilon +1)+10 \varepsilon +5\right)}{5 y  \varepsilon  (4 \varepsilon +1)} \tilde{G}_{1,\underline{1},1,0,0,1,\underline{1},\underline{1}, 1} \notag \\
& \quad -\frac{(3 y -5) \left(y ^2-1\right) (\varepsilon^2 -1)}{10 y  \varepsilon  \left(8 \varepsilon ^2+6 \varepsilon +1\right)} \tilde{G}_{1,\underline{1},2,0,0,1,\underline{1},\underline{1}, 1} \, . \notag
	\end{align}
\endgroup
The non planar MIs satisfy \eref{eq:diffnonplanar} where $\tilde{A}_{\pm 1}$ are
\begin{equation}
\!\!\!\!\!\!\!\!\!
	\begin{aligned}
		\tilde{A}_{+ 1}  = \left(
\begin{array}{cccccc}
 -2 & 0 & 0 & 0 & 0 & 0 \\
 -4 & -2 & 0 & 0 & 0 & 0 \\
 -8 & 0 & -2 & 0 & 0 & 0 \\
 8 & 0 & 0 & -2 & 0 & 0 \\[3pt]
 \frac{8048}{25} & \frac{4832}{75} & -\frac{5036}{75} & 24 & -\frac{27}{5} & -\frac{68}{15} \\[3pt]
 -\frac{11056}{25} & -\frac{2168}{25} & \frac{2264}{25} & -36 & \frac{24}{5} & \frac{22}{5} \\
\end{array}
\right) , \,
\tilde{A}_{- 1}  = \left(
\begin{array}{cccccc}
 0 & 0 & 0 & 0 & 0 & 0 \\
 4 & 1 & 0 & 0 & 0 & 0 \\
 0 & 0 & 2 & 0 & 0 & 0 \\
 -8 & 0 & 0 & -2 & 0 & 0 \\[3pt]
 -\frac{17872}{75} & -\frac{1088}{25} & \frac{1156}{25} & -24 & -\frac{43}{5} & -\frac{24}{5} \\[3pt] \frac{8368}{25} & \frac{1496}{25} & -\frac{1592}{25} & 36 & \frac{66}{5} & \frac{38}{5} \\
\end{array}
\right) \, .
	\end{aligned}
\end{equation}


\addcontentsline{toc}{section}{References} 
{\footnotesize \bibliography{eikonal} }
\bibliographystyle{utphys}

\end{document}